\documentclass{revtex4-1}%

 \usepackage{graphics}
\usepackage{graphicx}
\usepackage{epsfig}
\usepackage{amssymb}
\usepackage{xcolor}
\usepackage{color}
\usepackage{indentfirst}
\usepackage[utf8]{inputenc}
\usepackage{amsmath}
\usepackage{mathrsfs}

\begin{document}
    \title{Electromagnetic dipole moment and time reversal invariance violating interactions
    for high energy short-lived particles  in bent and straight crystals at Large Hadron Collider}
    \author{V.G. Baryshevsky }

    \begin{abstract}{
The channelled particle, which moves in a crystal, alongside with
electromagnetic interaction also experiences weak interaction with
electrons and nuclei, as well as strong interaction with nuclei.
Measurements of polarization vector and angular distribution of
particles scattered by axes (planes) of unbent crystal enable to
obtain limits for the EDM value and for values of constants
describing P- and T-odd interactions.
The same measurements also allow studying magnetic dipole moment
of charged and neutral particles.
Investigation of left-right asymmetry by the use of two unbent
crystals makes it possible to measure EDM, MDM and other constants
without studying the angular distribution of decay products of
scattered particles:
it is sufficient to measure the intensity of flow of particles
experienced double scattering.
Spin precession of channelled particles in bent crystals at the
LHC gives unique possibility for measurement of constants
determining $T_{odd}$, $ P_{odd}$ (CP) violating interactions and
$P_{odd}$, $T_{even}$ interactions of baryons with electrons and
nucleus (nucleons), similarly to the possibility of measuring
electric and magnetic moments of charm, beauty and strange charged
baryons.
Methods to separate P-noninvariant rotation from the MDM- and
EDM-induced ($T_{odd}$) spin rotations are discussed.
    }
\end{abstract} 
    %
    \maketitle

\section{Introduction}

The violation of parity (P) and time reversal (T) symmetries lead
to appearance of numerous processes allowing investigation of
physics beyond the Standard Model (SM). Recently, it has been
proposed to search  for the electromagnetic dipole moments (EDM)
of charged short lived heavy baryons using bent crystals at LHC
\cite{bn11,b10}. According to \cite{b29,b33} bent and straight
crystals allow investigation of both EDM of charged and neutral
baryons,and $ P_{odd} T_{even} $ and $ P_{odd} T_{odd} $
interactions between short lived baryons and electrons and nuclei.

This paper is devoted to consideration of influence of P- and
T-odd effects, which accompany interaction of charged and neutral
baryons with electrons and nuclei in crystals, on EDM measurement
at the LHC.
Experimental methods to distinguish different contributions to
spin rotation are suggested.

\section{Relativistic particles spin interactions with crystals}
Owing to quasiclassical nature of high energy particle movement in
crystal, to describe evolution of its spin in electromagnetic
fields inside of the crystal,Thomas--Bargmann--Michel--Telegdi
(T-BMT) equations \cite{bn14} are used. Let us consider a particle
with spin $S$ which  moves in the electromagnetic field.
The term ''particle spin'' here means the expected value of the
quantum mechanical spin operator $\hat{\vec{S}}$
(hereinafter the symbol marked with "hat"
means a quantum mechanical operator).
Movement of  high-energy particles (Lorentz-factor $ \gamma\gg 1
$) in non-magnetic crystals (magnetic field $ \vec{B}=0$) will be
considered below.

In this case spin motion is described by the
Thomas--Bargmann--Michel--Telegdi equation (T-BMT)  as follows:
\begin{equation}
\frac{d \vec{S}}{d t}=[\vec{S}\times\vec{\Omega}]\,,
\label{eq1}
\end{equation}
\begin{equation}
\vec{\Omega}=-\frac{e (g-2)}{2 mc}\left[\vec{\beta}\times\vec{E} \right],
\label{eq2}
\end{equation}
where $\vec{S}$ is the spin vector, $t$ is the time in the
laboratory frame, $m$ is the mass of the particle, $e$ is its
charge,  $\gamma$ is the Lorentz-factor, $\vec{\beta}=\vec{v}/c$,
where $\vec{v}$ denotes the particle velocity, $\vec{E}$ is the
electric field at the point of particle location in the laboratory
frame, $g$ is the gyromagnetic ratio (by definition, the particle
magnetic moment $\mu=(eg\hbar/2mc)S$, where $S$ is the particle
spin).

The T-BMT equation describes the spin motion in the rest frame of
the particle, wherein the spin is described by the three component
vector $\vec{S}$.
In practice the T-BMT equation well describes the spin precession
in the external electric and magnetic fields encountered in
typical present--day accelerators.
Study of the T-BMT equation enables one to determine the major
peculiarities of spin motion in an external electromagnetic field,
{ to describe spin rotation effect for particles in a crystal and
to apply it for measuring magnetic moments of unstable particles
\cite{b10,b1,b12,b8,b6,b2,b3,b4,b5,b9}.}
However, it should be taken into account that particles in an
accelerator or a bent crystal have energy spread and move along
different orbits.
This necessitates one to average the spin--dependent parameters of
the particle over the phase space of the particle beam.
That is why one must always bear in mind the distinction between
the beam polarization $ \vec{\xi} $ and the spin vector $\vec{S}$.
A complete description of particle spin motion can be made by the
use of spin density matrices equation (in more details see
\cite{b12,b22}).
For the case of ultra relativistic baryons T-BMT equations
supplied with the term that is responsible for interaction between
particle EDM and electric field can be written as follows ($
\gamma \gg 1, \gamma$ is the Lorentz-factor)
\cite{b20,b16,bn11,b10}:
\begin{equation}
\frac{d \vec{\xi}}{dt}= [\vec{\xi} \times \vec{\Omega}_{magn}] + [\vec{\xi} \times \vec{\Omega}_{EDM}],
\label{eq3}
\end{equation}
where $\vec{\xi}$ is the  particle polarization vector,
$\vec{\Omega}_{magn}=-\frac{e(g-2)}{2mc} [\vec{n} \times
\vec{E}_{\perp}] $, $\vec{\Omega}_{EDM}=
\frac{2ed}{\hbar}\vec{E}_{\perp}$. $\vec{E}_{\perp}$ is an
electric field component perpendicular to the particle velocity $
\vec{V} $, $\vec{n} $ is the unit vector parallel to the velocity
$ \vec{V} $, $ ed $ is the electric dipole moment.

It would be recalled that the particle refractive index in matter formed by different scatterers has the form:
\begin{equation}
n=1+\frac{2\pi N }{k^{2}}f\left( 0\right)\,,
\label{eq5.1}
\end{equation}
where $N$ is the number of scatterers per $cm^3$ and $k$ is the wave number of the particle incident on the target,
$f(0)\equiv f_{aa} (\vec{k}' -\vec{k}=0)$ is the coherent elastic zero angle scattering amplitude.
In this scattering,  momentum of the scattered particle $\vec{p} '=\hbar\vec{k}'$ ( where $\vec{k}'$ is a wave vector) equals to initial momentum $\vec{p}=\hbar\vec{k}$. Atom (nucleus) that was in quantum state before interaction with the incident particle characterized by stationary wave function $\Phi_{a}$ will stay in the same quantum state after interaction with the incident particle.
If the energy of interaction between particle and a scatterer depends on spin of the particle, then scattering amplitude $\hat{f}(\vec{k}' -\vec{k})$ can also depend on spin. As a consequence refractive index $\hat{n}$ (symbol $\hat{}$ means that mentioned magnitude is an operator in spin space of a particle) depends on spin as well \cite{b12}.

If the matter is formed by different scatterers, then
\begin{equation}
n=1+\frac{2 \pi}{k^2}\sum_{j}N_j f_j (0),
\label{eq4}
\end{equation}
where $N_j$ is the number of j-type scatterers per $cm^3$,  $f_j (0)$ is the amplitude of the particle coherent elastic zero-angle scattering by j-type scatterer.

Let us consider a relativistic particle refraction on the
vacuum-medium boundary (see \cite{b12}). The wave number of the
particle in the vacuum is denoted $k$. The wave number of the
particle in the medium is $\vec{k}'=kn$. As is evident the
particle momentum in the vacuum $\vec{p}=\hbar\vec{k}$ is not
equal to the particle momentum in the medium. Therefore, the
particle energy in the vacuum $E=\sqrt{\hbar^2 k^2 c^2 +m^2 c^4}$
is not equal to the particle energy in the medium
$E_{med}=\sqrt{\hbar^2 k^2 n^2 c^2 +m^2 c^4}$.

The energy conservation law immediately requires the particle in
the medium to have the effective potential energy $U_{eff}$. This
energy can be easily found from relation:

\begin{equation}
E=E_{med}+U_{eff},
\label{eq5}
\end{equation}
i.e.
\begin{equation}
U_{eff}=E-E_{med}= -\frac{2 \pi \hbar^2}{m \gamma} N f(E,0)= (2
\pi)^3 NT_{aa} (\vec{k}' -\vec{k}=0),
\label{eq5a}
\end{equation}
\begin{equation}
f(E,0)=-(2 \pi)^2 \dfrac{E}{c^2 \hbar^2} T_{aa} (\vec{k}'
-\vec{k}=0) = -(2 \pi)^2 \frac{m \gamma}{\hbar^2}T_{aa} (\vec{k}'
-\vec{k}=0),
\label{eq5b}
\end{equation}
where $T_{aa} (\vec{k}' -\vec{k}=0)$ is the matrix element of the T-operator describing elastic coherent zero-angle scattering .

Let us remind that T-operator  is associated with scattering matrix $S$ \cite{bn13, bn15}:
\begin{equation}
S_{ba}=\delta_{ba} -2 \pi i \delta (E_{b}-E_{a}) T_{ba},
\label{eq6}
\end{equation}
where $E_a$ is the energy of scattered particles before the collision, $E_b$ - after the collision, matrix element $T_{ba}$ corresponds to states $ a $ and $ b $ that refer to the same energy.

For the matter formed by different scatterers effective potential energy can be written as:
\begin{equation}
U_{eff}=- \dfrac{2 \pi \hbar^2}{m \gamma} \sum_{j}N_j f_j (E,0).
\label{eq7}
\end{equation}
Due to periodic arrangement of atoms in a crystal the effective potential energy is a periodic function of coordinates of a particle moving in a crystal \cite{b12}:

\begin{equation}
U(\vec{r})=\sum_{\vec{\tau}}U(\vec{\tau}) e^{i \vec{\tau} \vec{r}},
\label{eq8}
\end{equation}
where $\vec{\tau}$ is the reciprocal lattice vector of the crystal;

\begin{equation}
U(\vec{\tau})=\dfrac{1}{V}\sum_{j}U_j (\vec{\tau}) e^{i \vec{\tau} \vec{r_j}},
\label{eq9}
\end{equation}
here $V$ is the volume of the  crystal elementary cell,
$\vec{r_j}$ is the coordinate of the atom (nucleus) of type $ j $
in the crystal elementary cell.

\begin{equation}
U_{j}(\vec{\tau})=- \frac{2 \pi \hbar^2}{m \gamma} F_j (\vec{\tau}),
\label{eq10}
\end{equation}
According to (\ref{eq10}) effective potential energy
$U(\vec{\tau})$ is determined by amplitude $F_j
(\vec{\tau})=F_{jaa}(\vec{k}' -\vec{k}=\vec{\tau}).$ In contrast
to the case of chaotic matter where effective potential energy is
determined by the amplitude of elastic coherent scattering
$f(\vec{k}' -\vec{k})$, here it is defined by the amplitude $F
(\vec{\tau})$ (see Annex and \cite{b12}), which can be written as:
\begin{equation}
F_{j}(\vec{k}' -\vec{k})= f_{j}(\vec{k}' -\vec{k})-i\frac{k}{4 \pi}\int f_{j}^*(\vec{k}'' -\vec{k}')f_{j}(\vec{k}'' -\vec{k}) d\Omega_{ k''}.
\label{eq12}
\end{equation}
where $d\Omega_{ k''}$ means integration over all of the  vector $\vec{k}''$ directions,$|\vec{k}'|=|\vec{k}|=|\vec{k}''|$.

The occurrence of the amplitude $F(\vec{k}' -\vec{k})$ instead of
the amplitude of elastic coherent scattering  $f(\vec{k}'
-\vec{k})$ in (\ref{eq6}) is specified by the fact, that unlike of
an amorphous matter, the wave elastically scattered in a crystal,
due to rescattering by periodically located centers is involved in
formation of a coherent wave propagating through the crystal (see
Appendix \ref{ssec1} hereinafter).
\section{Effective potential energy of a spin-particle moving close to crystal planes (axes)}

Suppose a high energy particle moves in a crystal at a small angle
to the crystallographic planes (axes) close to the Lindhard angle.
This motion determined by the plane (axis) potential $ \vec{U}(x)
(\vec{U}(\vec{\rho}))$, which can be determined from
$\vec{U}(\vec{r})$ by averaging over distribution of atoms
(nuclei) in a crystal plane (axis).
Similar result is obtained when all the terms with $ \tau_y \neq 0, \tau_z \neq 0  $ for the case of planes  or $ \tau_z \neq 0 $ for the case of axes are removed from the sum (\ref{eq8}).

As a consequence for the potential of periodically placed axes we can write:

\begin{equation}
U(\vec{\rho})=\sum_{\vec{\tau}_{\perp}}U(\vec{\tau}_{\perp},\vec{\tau}_{z}=0) e^{i\vec{\tau}_{\perp}\vec{\rho}},
\label{eq172}
\end{equation}
$ z $-axis of the coordinate system is directed along the crystallographic axis. For the potential of a periodically placed planes we have:

\begin{equation}
U(x)=\sum_{\vec{\tau}_{x}}U(\tau_{x},\tau_{y}=0, \tau_{z}=0)
e^{i\tau_{x}x},
\label{eq182}
\end{equation}
$y,z $-planes of the coordinate system are parallel to the chosen
crystallographic planes family. Lets remind that according to
(\ref{eq9}-\ref{eq10}) the magnitude $U(\vec{\tau})$ is expressed
in terms of the amplitude $ F(\vec{\tau}) $.

Since the amplitude $ \hat{F}(\vec{k}'-\vec{k}) $ depends on spin,
the effective potential energy $\hat{U}$ depends on a spin as well
\cite{b12}. The magnitude $\hat{U}$ is the operator in spin space
of a particle incident on a crystal.

Lets express the amplitude $\hat{F}(q)$ as Fourier transformation of function $\hat{F} (\vec{r})$:
\begin{equation}
\hat{F}(\vec{q})=\int \hat{F}(\vec{r}') e^{-i \vec{q}\vec{r} '} d^3 r'.
\label{eq22}
\end{equation}
Considering mentioned above we can conduct summation of $\tau_x$ and $\vec{\tau_{\perp}}$ in (\ref{eq172},\ref{eq182}) using following expression:
\begin{equation}
\sum_{\tau_{x}} e^{i\tau x}=d_{x} \sum_{l} \delta(x-X_{l}),
\label{eq23}
\end{equation}
where $ d_x $ is the lattice period along axis $x$; $ X_l $ are coordinates of $ l $ plane.
\begin{equation}
\sum_{\tau_{x}, \tau_{y}} e^{i\vec{\tau}_{\perp}\vec{\rho}}=d_{x}
d_{y} \sum_{l} \delta(\vec{\rho}-\vec{\rho}_{l}), \label{eq24}
\end{equation}
where $\vec{\rho_l}$ is a coordinate of an axis, located in point $\vec{\rho_l}$; $d_x, d_y$ are lattice periods along axes $x$ and $y$.

As a result we obtain following expression for the effective potential energy of interaction between an incident particle and a plane (axis) (the lattice is assumed to consist of atoms of one kind):
\begin{eqnarray}
\hat{U}(x) &=& - \sum_{\tau_{x}}\frac{2 \pi \hbar^2 }{m \gamma V} \hat{F} (q_{x}=\tau_{x}, q_{y}=q_{z}=0) e^{i\tau_{x}x}= \nonumber\\
&=&-\frac{2 \pi \hbar^2 }{m \gamma V d_{y} d_{z}} \hat{F} (x, q_{y}=q_{z}=0),
\label{eq25}
\end{eqnarray}

\begin{eqnarray}
\hat{U}(\vec{\rho}) &=& -\frac{2 \pi \hbar^2 }{m \gamma V}
\sum_{\tau_{x}, \tau_{y}} \hat{F} (q_{x}=\tau_{x}, q_{y}=\tau_{y},
q_{z}=0)e^{i\vec{\tau}_{\perp}\vec{\rho}}= \nonumber\\
&=& -\frac{2 \pi \hbar^2 }{m\gamma d_z} \hat{F} (\vec{\rho}, q_{z}=0),
\label{eq26}
\end{eqnarray}
$d_z$ is the lattice period along the axis $z$.
\section{P-  and T-odd spin interactions in crystals}
Lets consider baryon scattering in thin crystal, when channeling
effects can be neglected. General expression for amplitude of
elastic coherent scattering of a spin $1/2$ particle by a spinless
(unpolarized) nuclei in presence of electromagnetic, strong and
$P$-, $T$-odd weak interactions can be written as:
\begin{equation}
\hat{F}(\vec{q})=A(\vec{q})+B(\vec{q}) \vec{\sigma}\vec{N}+B_{w}(\vec{q})\vec{\sigma}\vec{N_{w}}+B_{T}\vec{\sigma}\vec{N_{T}},
\label{eq32}
\end{equation}
where $ A(\vec{q}) $ is spin-independent part of scattering amplitude, which is caused by electromagnetic, strong and weak interactions.
$ \hbar\vec{q}=\hbar\vec{k}^{'} - \hbar\vec{k} $ is the
transmitted momentum, $\hbar\vec{k}^{'}$ is the momentum of the
scattered particle, $  \hbar\vec{k} $ is the momentum of the
incident baryon, $\vec{k}^{'},\vec{k}  $ are the wave vectors,
$\vec{N}=\frac{[\vec{k}^{'}\times\vec{k}]}{[\vec{k}^{'}\times\vec{k}]}
$, $\vec{N_{w}}=\frac{\vec{k}^{'}+\vec{k}}{|\vec{k}^{'}+\vec{k}|}
$, $\vec{N_{T}}=\frac{\vec{k}^{'}-\vec{k}}{|\vec{k}^{'}-\vec{k}|}
$, $\vec \sigma = (\sigma_x,\sigma_y,\sigma_z)$ are the Pauli
matrices.

The term, which is proportional to $ \vec{\sigma}\vec{N} $, is responsible for spin-orbit interaction contribution to scattering process. For electromagnetic interaction this contribution is determined by the particle magnetic moment. $ P_{odd} $$ T_{even} $ part of scattering amplitude, which is proportional to  $ \vec{\sigma}\vec{N_{W}} $ is determined by $ P_{odd} $$ T_{even} $ interactions of baryon with electron and nucleus. $ T_{odd} $ part of scattering amplitude, which is proportional to  $ \vec{\sigma}\vec{N_{T}} $ is determined by electric dipole moment and short range baryon-electron and baryon-nucleus $ T_{odd} $ interactions.

With amplitude $ \hat{F}(\vec{q}) $ one can find the  cross-section of particle scattering by a crystal and polarization vector of the scattered particle.
The scattering cross-section for a thin crystal can be written as
\cite{b8}:

\begin{equation}
\label{eq33}
\frac{d\sigma_{cr}}{d\Omega}=\frac{d\sigma}{d\Omega}\left\{(1-e^{-\overline{u^2}
    {q^2}}) +\frac{1}{N}\left|\sum_n e^{i\vec q\vec{r}_n^0}\right|^2 e^{-\overline{u^2} {q^2}}\right\},
\end{equation}
where $\vec{r}_n^0 $ is the coordinate of the center of gravity of the crystal  nucleus, $\overline{u^2}$ is the mean square  of thermal oscillations of nuclei in the crystal. The first term  describes incoherent scattering and the second one describes the coherent due to periodic arrangement of crystal nuclei (atoms). This contribution leads to the increase in the cross section $\sigma_{cr}$.

The value $ \frac{d\sigma}{d\Omega} $  describes baryon scattering cross section by atoms of crystal:
\begin{equation}
\label{eq34} \frac{d\sigma}{d\Omega}= tr \hat{\rho}
\hat{f^{+}}(\vec{q})\hat{f}(\vec{q}),
\end{equation}

\noindent where $ \hat{\rho} $ is the spin density matrix of the
incident particle.

The polarization vector of a particle that has undergone a single scattering event can be found using the following expression: 
\begin{equation}
\label{eq35} \vec\xi= \frac{\mbox{tr}\hat{\rho}
\hat{f^+}\vec\sigma \hat{f}}{\mbox{tr}\hat{\rho} \hat{f^+}
\hat{f}} = \frac{\mbox{tr}\hat{\rho} \hat{f^+} \vec\sigma
\hat{f}}{\frac{d\sigma}{d\Omega}}.
\end{equation}

Using (\ref{eq32}) and  (\ref{eq35}) one can obtain the following expressions for polarization vector of the scattered particle and differential cross-section:
\begin{equation}
\label{eq36}
\vec{\xi}=\vec{\xi_{so}}+\vec{\xi_{w}}+\vec{\xi_{T}},
\end{equation}
where $ \vec{\xi_{so}} $ is the change of polarization vector due to spin-orbit interaction,
$ \vec{\xi_{w}} $ is the change of polarization vector caused by weak parity violating   interaction, $ \vec{\xi_{T}} $ is the change of polarization vector caused by $T$-odd interaction.

\begin{eqnarray}
\label{eq371}
\vec{\xi_{so}} & = & \left\{(|{A}|^2 - |B|^2) \vec\xi_0 + 2
|B|^2 \vec N (\vec N\vec\xi_0)+ \right. \nonumber \\
&  & \left.  +  2 \texttt{Im} ({A}B^*)[\vec N\vec\xi_0] +2 \vec N \texttt{Re} ({A}B^*)\right\}\cdot \left(\frac{d\sigma}{d\Omega}\right)^{-1}, \nonumber \\
\vec{\xi_{w}}  & = &  \left\{(|{A}|^2 - |B_{w}|^2)
\vec\xi_0 + 2
|B_{w}|^2 \vec{N_{w}} (\vec{N_{w}}\vec\xi_0)+ \right. \nonumber \\
&  & \left.  + 2 \texttt{Im} ({A}B^*_{w})[\vec{N_{w}}
\vec\xi_0]+2 \vec{N_{w}} \texttt{Re}
({A}B^*_{w})\right\}\cdot
\left(\frac{d\sigma}{d\Omega}\right)^{-1}, \nonumber \\
\vec{\xi_{T}} & = & \left\{(|{A}|^2 - |B_{T}|^2)
\vec\xi_0 + 2
|B_{T}|^2 \vec{N_{T}}  (\vec{N_{T}} \vec\xi_0)+ \right. \nonumber \\
&  & \left.  +  2 \texttt{Im} ({A}B^*_{T})[\vec{N_{T}}
\vec\xi_0]+2 \vec{N_{T}}  \texttt{Re} ({A}B^*_{T})\right\}\cdot
\left(\frac{d\sigma}{d\Omega}\right)^{-1}.
\end{eqnarray}

\noindent The differential cross-section reads as follows:
\begin{eqnarray}
\label{eq381}
& & \frac{d\sigma}{d\Omega}=\mbox{tr}\rho f^+ f = \nonumber \\
& &   |{A}|^2 + |B|^2+|B_{w}|^2+|B_{T}|^2 +
2Re({A}B^*)\vec N \vec\xi_0 + \nonumber \\
& & 2Re({A}B_{w}^*)\vec{N_{w}} \vec\xi_0+
2Re({A}B_{T}^*)\vec {N_{T}} \vec\xi_0.
\end{eqnarray}

While deriving expressions (\ref{eq371}) and (\ref{eq381}) small
terms containing production of $ B_{w} $ and $ B_{T} $, which are
much smaller comparing to other terms,  were omitted.
\noindent According to (\ref{eq371}) the angle of polarization
vector rotation for a baryon scattered in a crystal is determined
by rotations around three mutually orthogonal directions (see
terms proportional to $ N$, $N_{w}$, $N_T $).
The indicated rotations are determined by electromagnetic, strong and weak $  P, T_{odd} $ interactions.
It should also be noted that initially unpolarized particle beam
($\xi_0=0 $) in a crystal acquires polarization
directed along one of three vectors $\vec N$, $\vec{N_{w}}$, $\vec
{N_{T}}$, which carries information about all types of
interaction too.
According to (\ref{eq381}) amplitudes interference results in
asymmetry in scattering caused by orientation of vectors $\vec
{N_{T}}$, $\vec N$, $\vec{N_{w}}$ with respect to $\vec\xi_0,
\vec{k}'$ and $ \vec{k}$ .
Therefore, the angular distribution of scattered particles
intensity is anisotropic.
Thus, measurements of the rotation angle and of the angular distribution of intensity for a
particle beam scattered by crystal axes enables to study $T_{odd }$
interactions of positive(negative) charged and neutral short-lived
baryons.
In particular, such measurements allow one to obtain restrictions
on electric dipole moment of short-lived baryons and other
constants describing these interactions.
The same makes also possible to study $P_{odd }$ and spin-orbit
interactions.
Let us remind that in case of electromagnetic interaction
spin-orbit interaction is determined by magnetic moment of the
particle.
{Important notice follows from (\ref{eq371}) and (\ref{eq381}):
measurement of the angular distribution (which depends on the spin
orientation) of products of the weak decay of the baryon is not a
single method to measure spin rotation angle for the short-lived
baryons.
The same can be made by studying left-right  asymmetry of the
baryon scattered by two crystals.
In this case it is sufficient to measure the angular distribution
of intensity of flow of double scattered particles.
The first crystal provides spin rotation, while the second enables
to study angular distribution of particles scattered by the first
crystal i.e. left-right asymmetry.
}
This can be realized, for example, by using set of crystals with
axes directed at small angle with respect to momentum of the
scattered particles and detection of either nuclear reaction or
ionizing losses inside crystal detector.
Computer modelling is essential for further analysis.
The angle of spin rotation $ \vartheta_{p} $ and scattering anisotropy  increases
significantly for a baryon in an unbent crystal, when particles
move at a small angle with respect to a crystal axes, since in
this case the scatterers density grows.
As a result, even for short-lived  beauty (bottom) baryons with
negative and neutral charge the $P_{odd }$ and $T_{odd }$
amplitude can be measured.
Study of the spin rotation of a particle, which
moves at a small angle with respect to a crystal axes, is hampered
by depolarization effect \cite{b7, b8, b12}.
Let us note that trajectories of the scattered particles, which
azimuth angles are in the vicinity  $ \varphi $ and $ \varphi +
\pi $ ($ z $ axes is directed along the crystal axes) contributes
to EDM (Todd interaction) spin rotation with opposite signs due to
different electric field signs. At the same time, the $P$-odd
rotation occurs around the momentum direction and is not affected
by the electric field direction.
As a result the $ T_{odd} $ spin rotation can be observed in unbent crystal if we use subtraction  of the measurements  results for angle ranges $
\varphi $ and $ \varphi + \pi $ from each other.
Such procedure leads to summation of contributions from $T$-odd
rotation.
Simultaneous measurement of spin orientation for all $ \varphi $
values (as well as for all polar angles for scattering by the
axes) provides intensity increase. For unbent crystals the same
measuring procedure  enables to use crystal with higher nucleus
charge that also contributes to effect increase.
The similar reasoning is valid for measuring the anomalous
magnetic moment by means of axial scattering in unbent crystals.
Subtraction procedure can be applied for measuring both anomalous
magnetic moment and EDM ( $ T_{odd} $ interactions) of neutral charm and beauty short-lived
baryons.

\section{Effective potential energy of a spin-particle moving close to the crystal planes (axes)}
\label{sec2} According to (\ref{eq25}) and(\ref{eq26}) to obtain
expression for potential energy of baryon in crystal one should
derive the expressions for contributions to $\hat{U}$ caused by
different types of interactions.

As was stated above elastic coherent scattering of a particle by an atom is caused by
electromagnetic interaction of the particle with the atom
electrons and nucleus as well as weak and strong nuclear
interaction  with electrons and nucleus.
The general expression for the amplitude of elastic scattering of
a particle with spin $\frac{1}{2}$  by a spinless or unpolarized
nuclei (\ref{eq32}) can be written as:

\begin{eqnarray}
&&\hat{F}(\vec{q})=A_{coul}(\vec{q})+A_{s}(\vec{q})+( B_{magn}(\vec{q}) + \nonumber \\
&&+B_{S}(\vec{q}))\vec{\sigma}[\vec{n}\times\vec{q}]+ \nonumber \\
&&+( B_{we}(\vec{q})  + B_{w nuc}(\vec{q}))\vec{\sigma}\vec{N}_{w} + \nonumber\\
&&+ (B_{EDM}(\vec{q}) + B_{Te}(\vec{q}) +
B_{Tnuc}(\vec{q}))\vec{\sigma}\vec{q},
\label{eq182a}
\end{eqnarray}
where $\vec{q}=\vec{k}'-\vec{k}, \vec{n}= \frac{\vec{k}}{k} ,
A_{coul} (\vec{q})$ is the spin-independent part of the amplitude
of elastic Coulomb scattering of a particle by an atom;  $A_{s}(\vec{q})$ is the
spin-independent part of the scattering amplitude, which is caused
by strong interaction (the similar contribution caused by weak
interaction it is negligibly small and hereinafter is omitted).

The spin-dependent amplitude, which is proportional to
$B_{magn}(\vec{q})$, is determined by electromagnetic spin-orbit
interaction.
The term proportional to $B_{s} (\vec{q})$ is responsible for the
contribution of  the spin-orbit strong interaction to a scattering
process of a baryon by a nucleus.

The term proportional to the parity odd pseudo scalar $
\vec{\sigma} \vec{N}_{w}$ (unit vector $\vec{N_w}=\frac{\vec{k}' +
    \vec{k}}{|\vec{k}'+\vec{k}|}$) includes two contributions:

\noindent a). Contribution to the amplitude proportional to
$B_{we}(q)$, which describes elastic scattering caused by the
parity violating weak interaction between the baryon and
electrons.

\noindent b). Contribution to the amplitude proportional to $
B_{wnuc}(\vec{q})$, which describes elastic scattering caused by
the parity violating weak interaction between the baryon and
the nucleus.

The term proportional to the time (T) violating (CP non-invariant)
pseudo scalar $ \vec{\sigma}\vec{q} $  includes three
contributions:

\noindent a). Contribution proportional to $ B_{EDM} (q) $
describes elastic scattering of the baryon with EDM by the atom's
Coulomb field.

\noindent b). Contribution proportional to $ B_{Te} (q) $
describes possible short-range T-non-invariant interaction between
the baryon and electrons.

\noindent c). Contribution to the amplitude, which is proportional
to $ B_{Tnuc} (q) $, describes scattering caused by
T-non-invariant interaction between the baryon and nucleons.

Using the amplitude $\hat{F}(q)$ potential energy $\hat{U}(\vec{r})$ can be expressed as a sum of terms that describe contribution of different interactions to $\hat{U}(\vec{r})$:

\begin{eqnarray}
\hat{U}(\vec{r})=
U_{coul}(\vec{r})+U_{S}(\vec{r})+\hat{U}_{magn}(\vec{r})+\hat{U}_{sp-orb}(\vec{r})
+ \hat{U}_{W}(\vec{r})+\hat{U}_{T}(\vec{r}), \label{eq333}
\end{eqnarray}
where
$U_{coul}(\vec{r})$ is Coulomb potential energy of interaction between baryon and crystal, which is detaily investigated during description of the effect of particles channeling in a crystal;
$U_{S}(\vec{r})$ describes spin independent contribution of nuclear interactions to the potential energy of interaction with crystal;
$\hat{U}_{magn}(\vec{r})$ describes contribution to $ \hat{U}(\vec{r}) $ caused by interaction between baryons magnetic moment and atoms electric field;
$\hat{U}_{sp-orb}(\vec{r})$ is contribution caused by spin-orbital nuclear interactions;
$ \hat{U}_{W}(\vec{r}) $ describes the contribution caused by parity violating weak interactions;
$\hat{U}_{T}(\vec{r})$ describes the contribution caused by T-violation interactions between baryon and crystal.

According to the analysis given in appendices of the paper
following expressions can be given  for potential energy
$\hat{U}(\vec r)$:

a) Potential energy for the plane $U_S (x)$:
\begin{equation}
U_{S} (x)= -\frac{2 \pi \hbar^2}{m \gamma d_y d_z} N_{nuc} (x)
A_{S} (0),
\label{eq31}
\end{equation}
where $N_{nuc} (x)= \iint N_{nuc} (x, y', z') dy' dz'$ is the
probability density for vibrating nuclei detection in point $x$
(in direction orthogonal to the chosen crystallographic plane).

Similarly, for the axis:
\begin{equation}
U (\vec{\rho})= -\frac{2 \pi \hbar^2}{m \gamma d_z} N_{nuc}
(\vec{\rho}) A_{S} (0), \label{eq322}
\end{equation}
where $N_{nuc} (\vec{\rho})= \int N_{nuc} (\vec{\rho}, z') dz'$,
$\vec \rho= (x,y)$ is a vector laying in a plane orthogonal to the
chosen crystallographic axis.

b) Effective potential energy determined by the anomalous magnetic moment
\begin{eqnarray}
\hat{U}_{magn}(x)&=&-\frac{e \hbar}{2 m c} \frac{g-2}{2} \vec{\sigma} [\vec{E}_{plane} \times \vec{n}]- \nonumber\\
&-&i \frac{1}{4 d_y d_z m c^2} (\frac{g-2}{2}) \frac{\partial}{\partial x}\overline{\delta V^2 (x)} \vec{\sigma}\vec{N}= \nonumber \\
&=&-(\alpha_m +i\delta_m)\vec{\sigma}\vec{N},
\label{eq41}
\end{eqnarray}
where
\begin{eqnarray}
&&\vec{N}= [\vec{n_x} \times \vec{n}],\nonumber \\
&&\alpha_m= \frac{e \hbar}{2 m c} \frac{g-2}{2} E_x, \nonumber \\
&&\delta_m= \frac{1}{4 d_y d_z m c^2} (\frac{g-2}{2})
\frac{\partial}{\partial x}\overline{\delta V^2 (x)},\nonumber
\end{eqnarray}
where $\overline {\delta V^2 (x)}\!=\!\int\!\left\{\overline{
\left[\int V_{coul} (x,y,z) dz \right]^2}\!-\!\left[ \overline {
\int V_{coul} (x,y,z) dz} \right]^2 \right\}\!dy$ is the
mean-square fluctuation of energy of Coulomb interaction between
baryon and atom.

Similarly for the case of axis it can be obtained:
\begin{equation}
\hat{U}_{magn}(\vec{\rho})= - \frac{e\hbar}{2mc}
\frac{g-2}{2}\vec{\sigma} [\vec{E}_{axis}\times \vec{n}]
+\hat{U}^{(2)}_{magn}(\vec{\rho}), \label{eq392}
\end{equation}
where
\begin{equation}
\hat{U}^{(2)}_{magn}(\vec{\rho})= -i \frac{1}{d_z m c^2}
(\frac{g-2}{2}) \vec{\sigma} [\nabla_{\rho} \overline{\delta V^2
    (\vec{\rho})} \times \vec{n}]. \nonumber
\label{eq39}
\end{equation}
For the axisymmetric case:
\begin{equation}
\hat{U}^{(2)}_{magn}(\rho)= -i \frac{1}{4 d_z m c^2}
(\frac{g-2}{2}) \frac{\partial}{\partial \rho}\overline{\delta V^2
    (\rho) } \vec{\sigma} [\vec{n}_{\rho}  \times \vec{n}], \label{eq40}
\end{equation}
\noindent where $\overline{\delta V^2 (\vec{\rho})} =  \overline{
    \left[ \int V_{coul} (\vec{\rho},z)dz \right]^2}  -
\left[\overline{\int V_{coul} (\vec{\rho},z)dz}\right]^2$ is the
mean-square fluctuation of energy of Coulomb interaction between
baryon and atom, $\vec{n}_{\rho}=\frac{\vec{\rho}}{\rho}$ is the
unit vector, $\vec{n}_{\rho} \perp \vec{n}$.

c) Effective potential energy $\hat{U}$ determined by spin-orbit interaction between baryon and nucleus:
\begin{equation}
\hat{U}_{s sp-orb}= -(\alpha_s +i\delta_s)\vec{\sigma} \vec{N},
\label{eq46}
\end{equation}
where in case of plane:
\begin{eqnarray}
&&\vec{N}= [\vec{n_x} \times \vec{n}],  \nonumber \\
&&\alpha_s=-\frac{2\pi \hbar^2}{m \gamma d_y d_z} \frac{\partial N_{nuc}}{\partial x} B'', \nonumber \\
&&\delta_s = \frac{2\pi \hbar^2}{m \gamma d_y d_z} B'
\frac{\partial N_{nuc}}{\partial x}.
\label{eq462}
\end{eqnarray}

In case of axis:
\begin{eqnarray}
\hat{U}_{s sp-orb} (\rho)=\frac{2 \pi \hbar^2}{m \gamma d_z}(B''
-iB')\vec{\sigma}[[\vec{\nabla_{\rho}} N_{nuc}(\vec{\rho}) \times
\vec{n}]. \label{eq463}
\end{eqnarray}

d) Effective potential energy $\hat{U}$ determined by  $ P_{odd} $ and $ T_{even} $ interactions
in the case of plane:
\begin{equation}
\hat{U}_{w}(x)= \hat{U}_{we}(x) + \hat{U}_{wnuc}(x)= - (\alpha_w (x) + i \delta_w (x)) \vec{\sigma}\vec{n},
\label{eq606}
\end{equation}
where
\begin{eqnarray}
\alpha_w (x)&=& \frac{2 \pi \hbar^2}{m \gamma d_y d_z} (\tilde{B}'_{we}(0) N_{e}(x) + \tilde{B}'_{wnuc}(0) N_{nuc}(x)), \nonumber\\
\delta_w (x)&=& \frac{2 \pi \hbar^2}{m \gamma d_y d_z} (\tilde{B}''_{we}(0) N_{e}(x) + \tilde{B}''_{wnuc}(0) N_{nuc}(x)).
\label{eq608}
\end{eqnarray}

In the case of axis:
\begin{eqnarray}
& & \hat{U}_{we}(\vec{\rho})\!=\!-\frac{2 \pi \hbar^2}{m \gamma d_z} \{\tilde{B}_{we}(0) N_{e}(\vec{\rho}) +\tilde{B}_{wnuc}(0) N_{nuc}(\vec{\rho})\}\vec{\sigma}\vec{N}_{w}, \nonumber \\
& & N_{e(nuc)}(\vec{\rho})\!=\!\int N_{e(nuc)} (\vec{\rho}, z)dz,
\label{eq605}
\end{eqnarray}
where $ \tilde{B}' $ and $ \tilde{B}'' $  are the real and
imaginary parts of $ \tilde{B} $, respectively.

e) T-violation interactions leads to the following contribution to potential energy:
\begin{equation}
\hat{U}_{T} (x)= \hat{U}_{EDM} + \hat{U}_{Te} + \hat{U}_{T nuc} = -(\alpha_{T}(x) + i\delta_{T}(x))\vec{\sigma}\vec{N}_{T},
\label{eq52}
\end{equation}
where $\alpha_{T} =\alpha_{EDM}+ \alpha_{Te}+ \alpha_{T nuc}, \delta_{T}=\delta_{EDM}+ \delta_{Te}+ \delta_{T nuc}$.

Energy of interaction between electric dipole moment with atoms electric field:

\begin{equation}
\hat{U}_{EDM} = -ed E_{pl}(x)\vec{\sigma}\vec{N}_{T}- i\frac{d}{2 d_y d_z \hbar c}\frac{\partial}{\partial x} \overline{\delta V^2 (x)} \vec{\sigma}\vec{N}_{T},
\label{eq49}
\end{equation}
where unit vector $\vec{N}_T$ is orthogonal to the plane,
$\vec{E}_{pl}(x)=E_x \vec{N}_{T}$.

Evidently, $\hat{U}_{EDM}$ can be expressed as:

\begin{equation}
\hat{U}_{EDM} = -(\alpha_{EDM} +i\delta_{EDM}) \vec{\sigma}\vec{N}_{T}.
\label{eq50}
\end{equation}
Similar to $\hat{U}_{magn}$ energy $\hat{U}_{EDM}$ has non-zero
both real and imaginary parts.
The expression for  $\hat{U}_{magn}$ converts to $\hat{U}_{EDM}$
when using replacement $ \frac{g-2}{2} \rightarrow
2\frac{d}{\lambda_c}$ ($ \lambda_c=\frac{\hbar}{mc} $ is the
Compton wave-length of the particle) and $\vec{N} \rightarrow
\vec{N}_T$.
Therefore:
\begin{equation}
\frac{U_{EDM}}{U_{magn}}=\frac{4d}{\lambda_c
(g-2)}=\frac{ed}{\mu_A}=\frac{D}{\mu_A}. \label{eq51}
\end{equation}
 Short range $ T_{odd}$ interactions give contributions to the
effective potential energy that can be written as:
\begin{equation}
\hat{U}_{Te(nuc)}(x)=-(\alpha_{Te(nuc)}+i\delta_{Te(nuc)})\vec{\sigma}\vec{N}_{T},
\label{eq55}
\end{equation}
where
\begin{eqnarray}
\alpha_{Te(nuc)}=\frac{2 \pi \hbar^2}{m \gamma d_y d_z} \tilde{B}''_{Te(nuc)} \frac{dN_{e(nuc)}(x)}{dx},\nonumber\\
\delta_{Te(nuc)}=\frac{2 \pi \hbar^2}{m \gamma d_y d_z}
\tilde{B}'_{Te(nuc)} \frac{dN_{e(nuc)}(x)}{dx}.\nonumber
\end{eqnarray}
 In the case of axis:
\begin{eqnarray}
\hat{U}_{Te(nuc)}(\vec{\rho})=-\frac{2 \pi \hbar^2}{m \gamma d_z}
(\tilde{B}_{Te(nuc)}'' + i \tilde{B}_{Te(nuc)}')\vec{\sigma}
\vec{\nabla}_{\rho} N_{Te(nuc)}(\vec{\rho}).
\end{eqnarray}

Thus in the experiment aimed to obtain the limit for the EDM
value, the limits for the scattering amplitude, which is
determined by short-range T(CP)-noninvariant interactions between baryons and
electrons, and nuclei, will be obtained as well.
The obtained values of these amplitudes for different interaction
types allows one to restore the values of corresponding constants,
too.
The simplest model for such a potential is the Yukawa potential
\cite{b28}.

Using it all the equations for $\alpha_{Te(nuc)}$ can be obtained
by replacement in (\ref{eq371}) of $ V_{coul}+V_{EDM} $ by
$V_{coul}+V_{T}$, $V_{T}=-d_{T} \vec{\sigma} \vec{r} \frac{e^
    {-\varkappa_{T}r}} {r^2}$.
Here $d_{T}$ is the interaction constant, $\varkappa_{T} \sim
\frac{1}{M_{T}} $, where $ M_{T} $ is the mass of heavy particles,
exchange of which leads to  interaction $V_{T}$ \cite{b28}.
It should be noted that constant $d_{T}$ for interaction between
the heavy baryon and the nucleon can be greater than that for
nucleon-nucleon interaction.
This effect can be explained by the reasoning similar to that
explaining expected EDM growth for the heavy baryon.
$ T_{odd} $ interaction mixes heavy baryon stationary states with different
parity more effectively that for light baryons due to probably smaller
spacing between energy levels corresponding to these states.

Lets note that as scattering amplitude $\hat{F}$ is a complex
value potential energy $ \hat{U} $ is also a complex value. The
real part of this energy describes changes in particles energy as
a result of interaction with the matter, the imaginary part
describes absorption.

Every spin dependent contribution to $\hat{U}$ is of following structure:
\begin{eqnarray}
\hat{A}=-(\alpha + i\beta)\vec{\sigma}\vec{N}. \label{eq5678}
\end{eqnarray}
Lets compare this expression with the one for energy of interaction between magnetic moment $ \vec{\mu} $ and magnetic moment $ \vec{B} $:
\begin{eqnarray}
\hat{U}_{magn}=-\vec{\mu}\vec{\sigma}\vec{B}. \label{eq5679}
\end{eqnarray}

It can be seen that terms proportional $\alpha$ in $\hat{A}$
causes spin rotation around $ \vec{N} $. The imaginary part shows
that absorption in matter depends on spin orientation regarding $
\vec{N} $. As a result spin component directed along $ \vec{N}$
can appear (spin dichroism occurs \cite{b12}).

The analogy between (\ref{eq5678}) and (\ref{eq5679}) leads us to
the conclusion that particle in the matter is affected by
pseudomagnetic fields caused by  strong and weak interactions ( in
neutron low energy area effects determined by such fields are
discovered and have been investigated for many years, see
\cite{b12}).
     %
Let now baryons with the polarization vector oriented at a certain
angle to the direction of $\vec{n}$ be incident on medium. This
baryon state can be considered as a superposition of two states
with polarizations along and opposite to the momentum direction
defined by unit vector $\vec{n}$. The wave function of a particle
before entering the target has the form
\begin{equation}
\label{6.14} \psi(\vec{r})= e ^{i\vec{k}\vec{r}}\chi_{n},
\chi_{n}=\begin{pmatrix}
c_{1}\\
c_{2}
\end{pmatrix}\,,
\end{equation}
or
\begin{equation}
\label{6.15} \psi(\vec{r})=c_{1} e
^{i\vec{k}\vec{r}}\begin{pmatrix}
1\\
0
\end{pmatrix}+ c_{2}e ^{i\vec{k}\vec{r}}\begin{pmatrix}
0\\
1
\end{pmatrix}\,,
\end{equation}
Choose the direction  $\vec{n}$  as the $z$-axis. The state of the type
$\left(\begin{smallmatrix}
1\\
0
\end{smallmatrix}\right)$  has the refractive index $n_{+}$, while the state of the type $\left(\begin{smallmatrix}
0\\
1
\end{smallmatrix}\right)$  has the refractive index $n_{-}$.
If a baryon with spin parallel to  vector $\vec{n}$ (spin state
$\left(\begin{smallmatrix}
1\\
0
\end{smallmatrix}\right)$) were incident on the target, its motion in matter
would be described by the wave function $\psi_{+}(r)=e^{ikn_{+}z}\left(\begin{smallmatrix}
1\\
0
\end{smallmatrix}\right)$.
If a baryon with spin antiparallel to $\vec{n}$ (spin state $\left(\begin{smallmatrix}
0\\
1
\end{smallmatrix}\right)$) is incident on the target, then in matter it is described by the wave
function ${\psi_{-}(r)= e^{ikn_{-}z}\left(\begin{smallmatrix}
    0\\
    1
    \end{smallmatrix}\right)}$. If a baryon with an arbitrary spin direction falls on the target,
    its wave function (see (\ref{6.15})) is the superposition of states $\left(\begin{smallmatrix}
1\\
0
\end{smallmatrix}\right)$ and $\left(\begin{smallmatrix}
0\\
1
\end{smallmatrix}\right)$. As a consequence, the wave function of baryons in matter is also the superposition of these states, and can be written as
\begin{equation}
\label{6.16} \psi(\vec{r})=\begin{pmatrix}
c_{1}& \psi_{+}& (\vec{r})\\
c_{2}& \psi_{-}& (\vec{r})
\end{pmatrix}= c_{1}e^{ikn_{+}z}\begin{pmatrix}
1\\
0
\end{pmatrix}+ c_{2}e^{ikn_{-}z}\begin{pmatrix}
0\\
1
\end{pmatrix}\,.
\end{equation}
Now let us consider how the polarization of baryons  changes as they  penetrate into the
interior of the target (with the growth of the target thickness). Suppose we have a detector
that transmits the particles with spin polarized along a certain direction in the detector
(the axis of the detector) and absorbs the particles with the opposite spin direction.
Such a detector is the analog of the Nicol prism \cite{vfel_Born} used in optics for analyzing
the polarization of light.
When polarized light is incident on the Nicol prism, one
component of the light polarization passes through it, while the component orthogonal
to the axis of the Nicol prism is absorbed.
In the case of baryons, a target with polarized nuclei may act as a detector.
As the scattering cross section of a polarized baryon depends on whether the
baryon spin is oriented along the direction of the polarization vector of the
nucleus or opposite to it, baryon absorption in the  detector exhibits the same
dependence.
Suppose that the axis of the detector is parallel to the $z$-axis
along which the detector nuclei are polarized. In this case the
detector analyzes those components of the baryon spin, which are
directed along the $z$-axis and opposite to it. From (\ref{6.16})
follows that the probability amplitude $A^{(+)}$ of finding the
baryon with spin state $\left(
\begin{smallmatrix}
1 \\
0 \\
\end{smallmatrix}
\right)$,
i.e., of finding the baryon polarized parallel to the $z$-axis, is given by the expression:
\[
A^{(+)}= (1\quad 0)\psi=c_1 e^{ikn_{+}z}\,.
\]
The probability
\begin{equation}
P^{(+)}_{z}=|(1\quad
0)\psi|^{2}=|c_{1}^{2}|e^{-2k\texttt{Im}n_{+}z}=|c_{1}|^{2}e^{-\rho\sigma_{+}z}\,.
\label{eq:probability+}
\end{equation}
Similarly, the probability $P^{(-)}_{z}$ of finding the baryon
polarized opposite to the $z$-axis is
\begin{equation}
P_{z}^{(-)}=|(0\quad
1)\psi|^{2}=|c_{2}|^{2}e^{-2k\texttt{Im}n_{-}z}=|c_{2}|^{2}e^{-\rho\sigma_{-}z}\,,
\label{eq:probability-}
\end{equation}
where $\sigma_{\pm}$ is the total cross section of scattering by
the nucleus of the baryon polarized parallel (antiparallel) to the
baryon momentum.
Since in matter $\texttt{Im}n_{+}\neq \texttt{Im} n_{-}$
($\sigma_{+}\neq \sigma_{-}$) owing to P-violation, one of the components of the
baryon spin wave function decays faster and at some depth the
rapidly damped component may be neglected. The beam will appear
polarized along the $z$-axis (along the direction of the
particle momentum $ n $).
Let us now rotate the detector so that its polarization axis
becomes perpendicular to the direction of $ n $. Choose the
direction of the polarization axis of the detector as the
$x$-axis.
Now the detector analyzes
those components of the baryon spin, which are directed along and opposite to the $x$-axis.
To determine the probability $P_{x}^{(\pm)}$ of finding the
component of the baryon spin parallel (antiparallel) to the
direction of the $x$-axis, one should expand the wave function

(\ref{6.16}) in terms of the spin wave functions $\chi_{x}^{\pm}$,
which are the eigenfunctions of operator $\hat{S}_{x}$ of the spin
projection onto the $x$-axis. They have the form
\[
\chi_{x}^{\pm}=\frac{1}{\sqrt{2}}
\begin{pmatrix}
\phantom{+}1 \\
\pm 1 \\
\end{pmatrix}\,.
\]
As a result, we find that the probabilities $P_{x}^{(\pm)}$ of
baryon spin polarization along and opposite to the $x$-axis change
with $z$ as:
\begin{eqnarray}
\label{ins}
P_{x}^{(+)}=\frac{1}{2}\left\{|c_{2}|^{2}e^{-2k\texttt{Im}n_{+}z}+|c_{2}|^{2}e^{-2k\texttt{Im}n_{-}z}\right.\nonumber\\
\left.+2|c_{1}|c_{2}|e^{-k\texttt{Im}(n_{+}+n_{-})z}\cos\left[k\texttt{Re}(n_{+}-n_{-})z+\delta\right]\right\}\,,\nonumber\\
P_{x}^{(-)}=\frac{1}{2}\left\{|c_{1}|^{2}e^{-2k\texttt{Im}n_{+}z}+|c_{2}|^{2}e^{-2k\texttt{Im}n_{-}z}\right.\nonumber\\
\left.-2|c_{1}|c_{2}|e^{-k\texttt{Im}(n_{+}+n_{-})z}\cos\left[k\texttt{Re}(n_{+}-n_{-})z+\delta\right]\right\}\,,
\end{eqnarray}
where $\delta=\delta_{1}-\delta_{2}$ is the difference of the initial phases of states with baryon spin polarization  along and opposite to the $z$-axis ($c_{1}=|c_{1}|e^{i\delta_{1}}$; $c_{2}=|c_{2}|e^{i\delta_{2}}$).
If at $z=0$, the baryon is polarized along $x$, i.e.,
\[
c_{1}=c_{2}=\frac{1}{\sqrt{2}}\,, \quad \delta=0\,,
\]
then with growing $z$ the polarization opposite to $x$ appears and further change of the polarization  acquires the character of oscillations.
As the baryons pass through the target, one of the components decays more strongly and the baryon beam eventually becomes polarized along or opposite to the $z$-axis. When a beam polarized along the $z$-axis is incident onto the target, no oscillations emerge: only damping occurs.
Using (\ref{6.16}), one can find the baryon polarization vector
\begin{equation}
\label{6.17} \vec{p}_{n}= \frac{\langle\psi|\vec{\sigma}|\psi\rangle}{\langle\psi|\psi\rangle}\,.
\end{equation}
As a result,
\begin{eqnarray}
\label{6.18} p_{nx}&=&
2\texttt{Re} c_{1}^{*}c_{2}\psi_{+}^{*}\psi_{-}\langle\psi|\psi\rangle^{-1}\,
\nonumber \\
p_{ny}&=&2 \texttt{Im} c_{1}^{*}c_{2}\psi_{+}^{*}\psi_{-}\langle\psi|\psi\rangle^{-1}\,,
\nonumber \\
p_{nz}&=&(|c_{1}\psi_{+}|^{2}-|c_{2}\psi_{-}|^{2})\langle\psi|\psi\rangle^{-1}\,.
\end{eqnarray}
Suppose that baryon spin in a vacuum is directed perpendicular to the
polarization vector of the nuclei. Choose this direction  as the
$x$-axis. In this case
\[
c_{1}=c_{2}=1/\sqrt{2}\,.
\]
Using relations \ref{6.18}, we obtain
\begin{eqnarray}
\label{6.19}
&&p_{nx}= \cos[k\texttt{Re}(n_{+}-n_{-})z]e^{-k\texttt{Im}(n_{+}+n_{-})z}\langle\psi|\psi\rangle^{-1}\,, \nonumber \\
&&p_{ny}=-\sin[k\texttt{Re}(n_{+}-n_{-})z]e^{-k\texttt{Im}(n_{+}+n_{-})z}\langle\psi|\psi\rangle^{-1}\,, \nonumber \\
&&p_{nz}=\frac{1}{2}(e^{-2k\texttt{Im}
    n_{+}z}-e^{2k\texttt{Im}n_{-}z})\langle\psi|\psi\rangle^{-1}\nonumber\\
&& p_{x}^{2}+p_{y}^{2}+p_{z}^{2}=1\,.
\end{eqnarray}
According to (\ref{6.19})\, as the baryon penetrates into the
interior of the target, its polarization vector rotates about the
particle's momentum direction $\vec n$ through the angle
\begin{equation}
\label{6.20} \theta=
k\texttt{Re}(n_{+}-n_{-})z=\frac{2\pi\rho}{k}\texttt{Re}(f_{+}-f_{-})z=\frac{Re(U_W^+ -U_W^-)}{\hbar}\frac{z}{c}.
\end{equation}

At the same time, as the baryons pass through matter, the
transverse components $p_{nx}$ and $p_{ny}$ of the polarization
vector  decay because baryon absorption depends on spin
orientation, and finally the beam appears to be polarized along or
opposite to the $z$-axis.
Thus, the dependence of the  absorption of baryons in the target
on the orientation of their spin results in the fact that the
polarization vector $\vec{p}_{n}$ (recall that $|\vec{p}_{n}|=1$)
not only rotates  about the $z$-axis (about the direction of
momentum) but also undergoes additional rotation in the direction
of the $z$-axis (the end point of the polarization vector moves
along the unit sphere).
If the dependence of absorption on spin orientation can be
neglected, the polarization vector rotates about the direction of
particle momentum $\vec{n}$ only in the $(x, y)$ plane.
In terms of kinematics, this phenomenon is analogous to the light polarization
plane rotation in a magnetic field (the Faraday effect), while spin oscillations along and opposite to the direction of the $x$-axis  are analogous to the transitions $K^{0}\rightleftarrows \bar{K}^{0}$ occurring in  regeneration of neutral $K$-mesons (see e. g. \cite{rins_85}).

Lets now consider particle moving in straight (non bent) crystal.
The expression for $ U $ contains group of terms proportional to
either electric field projection on to $x$ axis or derivative of
electrons and nuclei density  $\frac{dN_{e(nuc)}(x)}{dx}$. As a
result, particle moving between the planes experience  influence
of pseudomagnetic fields that reverse sign due to  transverse
oscillations of a channeled particle. This leads to the fact that
total spin rotation in such fields is suppressed (although
suppression fades with the growth of energy of the particle).
Mentioned above does not concern to spin rotation effect and  spin
dichroism caused by weak P-violation $ T_{even} $ interaction.This
effect increases with  the growth of crystal thickness (the effect
also occurs in amorphous medium) \cite{b12}.

\section{Relativistic particle spin rotation and interactions in bent crystals}
\begin{figure}[htbp]
    \centerline{\includegraphics[width=10 cm]{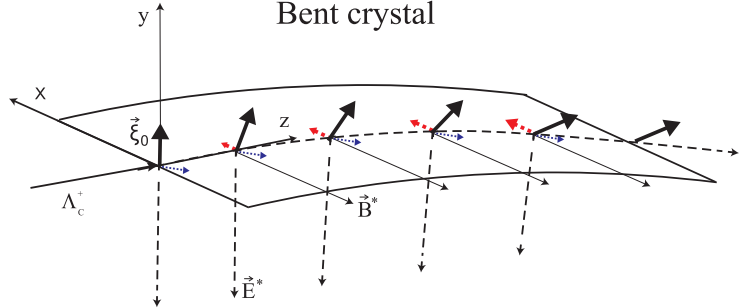}}
    \caption{Spin rotation caused by magnetic moment and T-reversal violation
        interactions (including EDM). Black arrows represent spin rotation about effective magnetic field
        (about bent axis, direction $ \vec{N_m} $), red arrows represent
        spin component caused by EDM (direction $ \vec{N}_T $),
        purple arrows represent the new effect - appearance of the spin component directed along  $ \vec{N_m}$ owing to the spin dichroism
        (spin rotation and dichroism in direction $ \vec{N}_T $ owing to T-reversal violation and P-violating interactions, is not shown here for simplicity). }
    \label{fig2}
\end{figure}

\begin{figure}[htbp]
    \centerline{\includegraphics[width=10 cm]{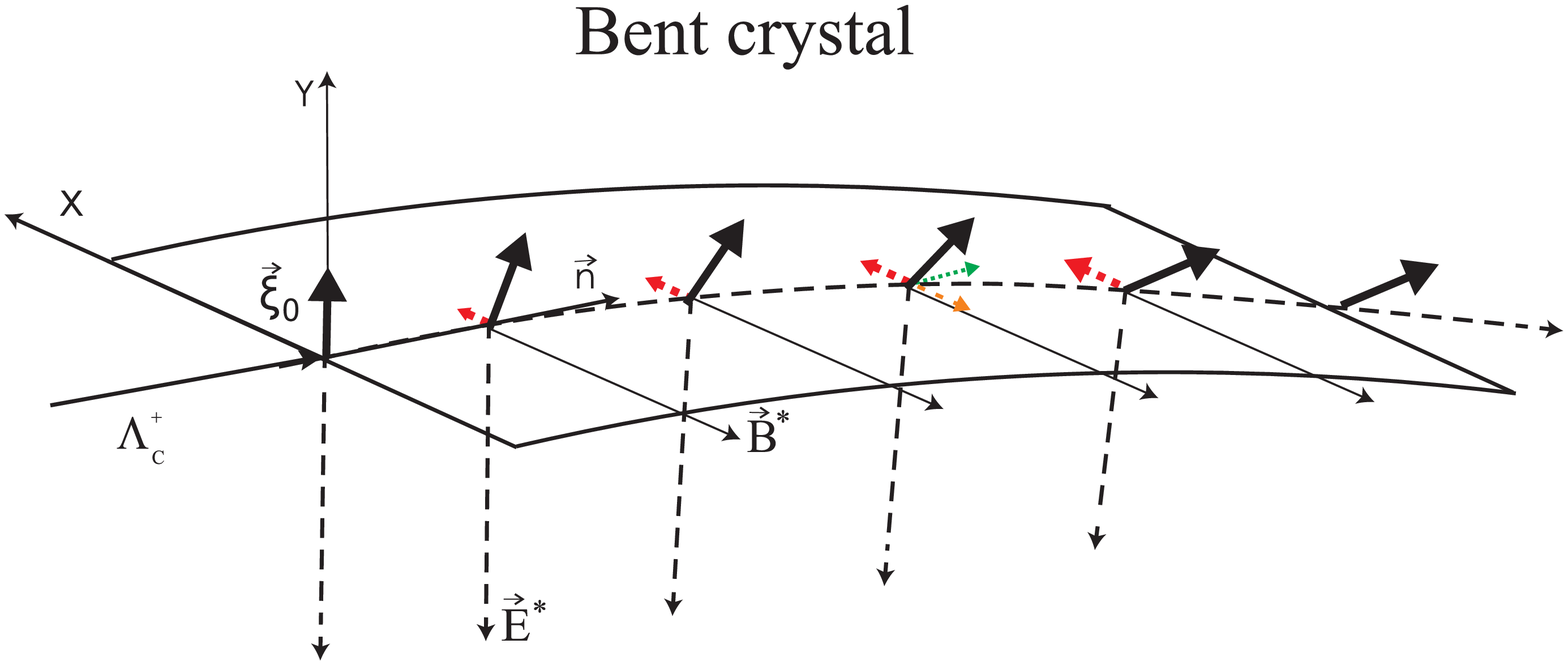}}
    \caption{Spin rotation caused by magnetic moment, T-reversal violation interactions
        (including EDM), P-violation spin rotation about direction $ \vec{n} $ (orange arrow) and spin component in direction $ \vec{n} $ caused by spin dichroism
        (green arrow).  Spin components caused by spin dichroism in direction $ \vec{N}_m $ and direction $ \vec{N}_T $ are not shown for simplicity.}
    \label{fig3}
\end{figure}

Lets now consider particle moving in bent crystal. { Expressions
for energy of interaction between a baryon and a crystal
plane(axis), which are obtained above, allow us to find the
equation describing evolution of the particle polarization vector
in a bent crystal.
The mentioned equations differ from those, which describe spin
evolution in external electromagnetic fields in vacuum, by
presence of terms, which define contributions from P and T(CP)
noninvariant interactions between electrons and nuclei to the spin
rotation.
Moreover, a new effect, which is caused by
nonelastic processes, arises: along with the spin precession
around vectors $\vec{N}_m$, $\vec{N_T}$, $\vec{n}$, the spin
components directed along vectors $\vec{N}_m$, $(\vec{N_T}$,
$\vec{n})$ appear and, thus, spin dichroism occurs.
Equations, which describe spin rotation in this case, can be
obtained by the following approach \cite{b12}.
Spin wave function $|\Psi(t)>$ meets the equation as follows: 
\begin{equation}
ih \frac{\partial |\Psi(t)>}{\partial t}= \hat{U}_{eff}|\Psi(t)>.
\label{eq56}
\end{equation}
Baryon polarization vector $\vec{\xi}$ can be found via $|\Psi(t)>$:
\begin{equation}
\vec{\xi}=\frac{<\Psi(t)|\vec{\sigma}|\Psi(t)> }{<\Psi(t)|\Psi(t)>},
\label{eq57}
\end{equation}
Thus the equation for spin rotation of a particle $(\gamma \gg1)$,
which moves in a bent crystal, reads as follows:
}
\begin{eqnarray}
\frac{d \vec{\xi}}{dt}&=&[\vec{\xi} \times \vec{\Omega}_{mso} ]-\frac{2}{\hbar} (\delta_m (x)+ \delta_{s0}(x))\{\vec{N}_m -\vec{\xi}(\vec{N}_m\vec{\xi})\} + \nonumber\\
&+& [\vec{\xi} \times \vec{\Omega}_{T}] + \frac{2}{\hbar}(\delta_{EDM}(x)+ \delta_{Te}(x)+ \delta_{Tnuc}(x)) \{\vec{N}_T -\vec{\xi}(\vec{N}_T\vec{\xi})\}+ \nonumber\\
&+& [\vec{\xi} \times \vec{\Omega}_{W}]-\frac{2}{\hbar}\delta_W \{\vec{n} -\vec{\xi}(\vec{n}\vec{\xi})\}
\label{eq**}
\end{eqnarray}
where
$\vec{\Omega}_{mso}=\vec{\Omega}_{MDM}+\vec{\Omega}_{so}=-\left(
\frac{e(g-2)}{2mc}E_x (x)+\frac{2}{\hbar}\alpha_{so}(x)\right)
\vec{N}_{m} $,
$\vec{\Omega}_{T}=\vec{\Omega}_{EDM}+\vec{\Omega}_{Ten}=\frac{2}{\hbar}(dE_x(x)+
\alpha_{Te}(x)+\alpha_{Tnuc}(x))\vec{N}_{T} $, $\vec{\Omega}_{w}
=\frac{2}{\hbar} \alpha_w \vec{n} $.
Let us note that vector $[\vec{n} \times \vec{E}]$ is parallel to
vector $\vec{N}_{m}=[\vec{n} \times \vec{n}_{x}]$ and $\vec{N}_{m}=-\vec{N}$ (see (\ref{eq46})), vector
$\vec{E}$ is parallel to $\vec{N}_{T}=\vec{n}_{x}$, $
\vec{n}=\frac{\vec{k}}{k} $ is the unit vector parallel to the
direction of the particles momentum.
Lets compare equations (\ref{eq3}) and (\ref{eq**}):

\begin{equation}
\frac{d \vec{\xi}}{dt}= [\vec{\xi} \times \vec{\Omega}_{magn}] + [\vec{\xi} \times \vec{\Omega}_{EDM}],
\label{eq3}
\end{equation}
%
%
%
According to (\ref{eq**}) baryon spin rotates around three axes
\cite{b29}: effective magnetic field direction
$\vec{N}_{m}||[\vec{n}\times\vec{E}]$, electric field direction
$\vec{N}_{T}||\vec{E}$ and momentum direction $\vec{n}$.
Nonelastic processes in crystals result in the new effect: terms
proportional to $\delta$ lead to appearance of the
polarization vector component in directions of vectors $\vec{N}_{m}$,
$\vec{N}_{T}$ and $\vec{n}$.
Let's pay attention to the fact that appearance of the spin
component directed along effective magnetic field $B^*$ ($
\vec{N}_m $ direction) is caused by both spin rotation around
direction of the electric field $\vec {E}$ ($\vec{N}_T$ direction)
due to T-noninvariant violation and spin dichroism due
to nonelastic processes at interaction between the magnetic moment
and the bent crystal. It can be seen that appearance of such spin
component imitates the result of the T-noninvariant rotation
(Figures \ref{fig2}, \ref{fig3}).
%
%
%
%
Contributions to equation (\ref{eq**}), which are caused by the
interaction between baryon and nuclei, depend on distribution of
nuclei density  $N_{nuc}(x)$ (see terms proportional to
$\alpha_{s0}(x), \delta_{s0}(x), \alpha_{Tnuc}(x),
\delta_{Tnuc}(x)$). As a result, for positively charged particles,
moving in the channel along the trajectories located in the center
of the channel, such contributions are suppressed.
Thus, according to (\ref{eq**}), when one conducts and interprets
experiments aimed for measuring EDM, one should consider the fact
that  measuring spin rotation provides information about the sum
of contributions to T-noninvariant rotation.
The mentioned rotation is determined  by both EDM and short-range
CP-noninvariant interactions. Nonelastic T-noninvariant processes
lead to spin dichroism in direction of $ \vec{N}_{T}$
as well, which gives additional opportunities for EDM measurement.
Let's evaluate the most important new effects described by the
equation(\ref{eq**}) and consider the contribution to spin
rotation caused by spin dichroism in direction of $
\vec{N}_{m} $.
According to (\ref{eq41}) coefficient $ \delta_{m} $ reads as
follows:
\begin{eqnarray}
\delta_{m}&=& \frac{1}{4d_yd_zmc^2} (\frac{g-2}{2}) \frac{\partial}{\partial x} \overline{\delta V^2 (x)}= \nonumber \\
&=&\frac{1}{4d_yd_z}mc^2(\frac{g-2}{2}) \frac{\partial}{\partial x} \int \left\{ \overline{\left[\int V_{coul} (x,y,z)dz\right]^2} - \right. \nonumber \\
&-& \left. \left[\int \overline{V_{coul} (x,y,z)}dz\right]^2\right\}dy,
\label{eq59}
\end{eqnarray}
where $V_{coul} (x,y,z)=\sum_i V_e (x-x_i, y-y_i,z-z_i) - V_{nuc}
(x-\eta_{fx},y-\eta_{fy}, z-\eta_{fz})$, $x_{i},y_{i},z_{i}$ are
the coordinates of the $i$-th electron in atom,
$ \eta_{fx}$, $\eta_{fy}$, $\eta_{fz}$ are the coordinates of the
{atom} nucleus.
Let us choose the position of equilibrium point for the
oscillating nucleus as the origin of coordinates.
The {overline denotes averaging of electrons' and nuclei'
    positions over electron density distribution and nuclei
    oscillations; in other words, averaging with wave-functions of
    atoms in crystal.}
By means of these functions, the density distribution can be
expressed as follows:
\begin{equation}
N(\vec{r}_1, \vec{r}_2......\vec{r}_z,\vec{\eta})=N_e (\vec{r}_1, \vec{r}_2......\vec{r}_z,\vec{\eta}) N_{nuc} (\vec{\eta}),
\label{eq60}
\end{equation}
where $N_{e}$ is the density distribution of electrons in atom,
$N_{nuc}(\vec{\eta})$ is the density distribution of nucleus
oscillations.
Let's introduce the function $W(x,y) = \int V(x,y,z)dz $. From
(\ref{eq59}) we have:
\begin{eqnarray}
&&W(x,y)=\sum_i \int V_e (x-x_i, y-y_i,\xi) d\xi - \nonumber \\
&-&\int V_{nuc} (x-\eta_x, y-\eta_y,\xi)d\xi = \nonumber \\
&=& \sum_i W_e (x-x_i, y-y_i)-\nonumber \\
&-&W_{nuc}(x-\eta_x, y-\eta_y)
\label{eq61}
\end{eqnarray}
\begin{eqnarray}
&&\overline{W^2 (x,y)} =\int [\sum_i W_e (\vec{\rho}-\vec{\rho}_i) - W_{nuc} (\vec{\rho}-\vec{\eta}_{\perp})]^2 \times \nonumber \\
&&\times N_e (\vec{\rho}_1-\vec{\eta}_1,.....\vec{\rho}_z-\vec{\eta}_{\perp}) \nonumber \\
&&N_{nuc}(\vec{\eta}_{\perp}) d^2\rho_1 d^2\rho_z d^2 \vec{\eta}_{\perp},
\label{eq62}
\end{eqnarray}
where $\vec{\rho}=(x,y),\vec{\eta}_{\perp} =(\eta_{x}, \eta_{y}), Z $ is the number of electrons in atom.
In other words:
\begin{eqnarray}
&&\overline{W^2 (\vec{\rho})} =\int \{ (\sum_i W_e (\vec{\rho}-\vec{\rho}_i))^2-\nonumber \\
&&- 2\sum W_e (\vec{\rho}-\vec{\rho}_i) W_{nuc} (\vec{\rho}-\vec{\eta}_{\perp}) + \nonumber \\
&&+ W^2_{nuc} (\vec{\rho}-\vec{\eta}_{\perp})\}N_e (\vec{\rho}_1-\vec{\eta}_1,........\vec{\rho}_z-\vec{\eta}_{\perp})\nonumber \\
&&N_{nuc}(\vec{\eta}_{\perp}) d^2\rho_1 d^2\rho_z d^2 \vec{\eta}_{\perp}. \nonumber \\
\label{eq63}
\end{eqnarray}
The result of averaging $\overline{W^{2}(\rho)}$ includes two
contributions:  that for density distribution of a single electron
in atom and one dependent on coordinates of two electrons in the
atom, which describes pair correlations in electrons positions in
the atom.
However, the influence of pair correlations will be ignored during
the estimations. As a result the expression (\ref{eq63}) can be
represented as follows:
\begin{eqnarray}
\overline{W^2 (\rho)} &=&\int d^2 \eta_{\perp} \{Z [\langle W_e^2(\vec{\rho},\vec{\eta}_{\perp}) \rangle _e -\nonumber \\
&-&\langle W_e (\rho,\eta_{\perp})\rangle _e^2] + Z^2 <W_e(\vec{\rho},\vec{\eta}_{\perp})> _e^2 - \nonumber \\
&-& 2Z <W_e(\vec{\rho},\vec{\eta}_{\perp})> W_{nuc}(\vec{\rho}-\vec{\eta}_{\perp}) + \nonumber \\
&+& W^2_{nuc}(\vec{\rho}-\vec{\eta}_{\perp}) \},
\label{eq64}
\end{eqnarray}
where the function
\begin{eqnarray}
<W_e(\vec{\rho},\vec{\eta}_{\perp})>_e= \int W_e (\vec{\rho}-\vec{\rho} ') N_e (\vec{\rho} '-\vec{\eta}_{\perp})d^2 \rho ' \nonumber\\
<W_e^2 (\rho,\vec{\eta}_{\perp} )>_e= \int W_e^2(\vec{\rho}-\vec{\rho}')N_e (\vec{\rho} '-\vec{\eta}_{\perp})d^2 \rho',\nonumber
\end{eqnarray}
that means
\begin{eqnarray}
& &\overline{W^2(\vec{\rho})}\!=\! \int\! d^2 \eta_{\perp}\! \left\{\! Z [<W_e^2(\vec{\rho},\vec{\eta}_{\perp})>_e\!-\!<W_e(\vec{\rho},\eta_{\perp})>^2_e] +\right. \nonumber \\
& & \left. + (Z<W_e(\vec{\rho},\vec{\eta}_{\perp})>_e -
W_{nuc}(\vec{\rho}-\vec{\eta}_{\perp}))^2
\right\}N_{nuc}(\vec{\eta}_{\perp}). \label{eq66}
\end{eqnarray}
According to (\ref{eq59}), the function $\int
\left[\overline{W^2(\vec{\rho})} -
\overline{W(\vec{\rho})}^2\right]dy$ determines the expression for
$ \delta_m $. It should be noted that when fluctuations caused by
nuclei oscillations are neglected, only fluctuations, which are
determined by distribution of electrons' coordinates in the atom,
are left.
As a result, the following equation for $ \delta_m $ can be obtained:
\begin{equation}
\delta_m= \frac{1}{4 d_y d_z} mc^2  \left(\frac{g-2}{2}\right) \frac{\partial}{\partial x} \int \left\{ \overline{W^2 (x,y)} - \overline{W (x,y)}^2 \right\} dy,
\label{eq67}
\end{equation}
where $$ \overline{W(\vec{\rho})}= \int \left\{Z \overline{W_e
    (\vec{\rho}, \eta_{\perp})}^e -W_{nuc} (\vec{\rho}- \eta_{\perp})
\right\}N_{nuc} (\eta_{\perp})d^2 \eta_{\perp},$$ $$\overline{W_e
    (\vec{\rho}, \eta_{\perp})}^e = \int W_e (\vec{\rho}-\vec{\rho}
')N_e (\vec{\rho} '-\vec{\eta_{\perp}})d^2 \rho ' .$$
Lets neglect nuclei oscillation to estimate the value $ \delta_m $.
In this case contribution to meansquare fluctuation of the energy of Coulomb interaction between baryon and atom is caused by fluctuations of positions of electrons in atom. As a result the expression for $ \delta_m $ takes the following form:
\begin{eqnarray}
\delta_m=\frac{1}{4 d_y d_z} mc^2\left(\frac{g-2}{2}\right)\times \nonumber \\
\times \frac{\partial}{\partial x} \int\left\{ z[<W_e^2(x,y)>_e-<W_e(x,y)>^2_e]\right\} dy
\label{eq777}
\end{eqnarray}
where function $ W_e(x,y)=\int V(x-x',y-y',z)dz $.
In expression (\ref{eq777}) averaging of electrons distribution in atom (see explanations to the expression (\ref{eq64}) ) is conducted over variables $x',y'$.
To conduct estimations, for  Coulomb energy of interactions
between baryon and electrons shielded Coulomb potential is used.
Lets suppose that electrons are distributed uniformly in area
about the same size as the radius of the shielding. In this case,
the following estimations for $ \delta_m $  can be obtained: $
\delta_m \sim 10^{8}\div10^{9}sec^{-1}$ depending on the position
of the  baryon trajectory in the planar channel. According to
\cite{b10, bn11} estimated experimental sensitivity for EDM is $ed
\sim 10^{-17}e $ $ cm $. Spin rotation frequency
$\Omega_{EDM}=\frac{2edE}{\hbar}$. The field $ E $ affecting
baryons in bent crystal can be obtained from the expression $
E=\frac{m\gamma c^2}{eR}$, where $ R $ is the radius of crystal
curvature.
Therefore $\Omega_{EDM} = 2\frac{d}{R} \frac{W}{\hbar}$, where $ W $ is the energy of baryon. For $ R=30m$,
$ d \sim 10^{-17}cm  $ and $ W=1TeV $ we have $ \Omega_{EDM} \simeq 10^7 sec^{-1} $
As a result, the nonelastic processes, which are caused by
magnetic moment scattering, can imitate the EDM contribution. Surely, more detailed computer simulation is needed.
The contributions of $ P_{odd} $  and $ T_{even} $ rotation effect to the general spin rotation can be evaluated by the following way.
Precession frequency $\Omega_{w}$ is determined by the real part of the amplitude of
baryon weak scattering by an electron (nucleus).
This amplitude can be evaluated by Fermi theory
\cite{rins_85,b26+} for the energies, which are necessary for W
and Z bosons production or smaller:
\begin{equation}
\label{eq29.1}
ReB\sim G_{F}k=10^{-5}\frac{1}{m^{2}_{p}}k=10^{-5}\frac{\hbar }{m_{p} c}\frac{m}{m_{p}\gamma}=10^{-5}\lambda_{cp} \frac{m}{m_{p}\gamma} ,
\end{equation}
where $ G_{F} $ is the Fermi constant, $m_{p}$ is the proton mass,
$\lambda_{cp} $ is the proton Compton wavelength. For particles
with energy from hundreds of GeV to TeV $ReB\sim G_{F}k=10^{-16}$
cm.
For different particle trajectories in a bent crystal  the value
of precession frequency $ \Omega_{w} $ could vary in the range $
\Omega_{w}\simeq 10^{3}\div 10^{4} sec^{-1}$.
Therefore, when a particle passes 10~cm in a crystal, its spin
undergoes additional rotation around momentum direction at angle $
\vartheta_{p} \simeq 10^{-6}\div 10^{-7}$ rad.
The effect grows for a heavy baryon as a result of the mechanism
similar to that of its EDM growth (see the explanation for the
growth of constant $d_T$ mentioned above).
Absorption caused by parity violating weak interaction also
contributes to change in spin direction (see the terms
proportional to $\delta_W$ in (\ref{eq**}) ).
This change is caused by the imaginary part of weak scattering
amplitude and is proportional to the difference of total
scattering cross-sections  $ \sigma_{\uparrow\uparrow} $ and $
\sigma_{\downarrow\uparrow} $ \cite{b29}.
This difference is proportional to the factor, which is determined
by interference of coulomb and weak interactions for baryon
scattering by an electron, as well as by interference of strong
(coulomb) and weak interactions for baryon scattering by nuclei
\cite{b29}.
\begin{equation}
\sigma_{\uparrow\uparrow (\downarrow\uparrow)}=\int| f_{c(nuc)}+B_{0w}\pm B_{w}|^{2} d\Omega ,
\label{eq30.1}
\end{equation}
\begin{equation}
\sigma_{\uparrow\uparrow} - \sigma_{\downarrow\uparrow}= 2\int[(f_{c(nuc)}+B_{0w})B^{*}+( f_{c(nuc)}+B_{0w})^{*}B]d\Omega .
\label{eq31.1}
\end{equation}
When baryon trajectory passes in the area, where collisions with
nuclei are important (this occurs in the vicinity
of potential barrier for positively charged particles),  the value
$ \delta_W \sim 10^{6}\div10^{7} sec^{-1}$.
Similar to the real part $ReB$ for the case of heavy baryons the
difference in cross-sections grows.
Multiple scattering also
contributes to spin rotation and depolarization \cite{b8,b12,b22,b29}.
Particularly,  due to interference of magnetic, weak and coulomb
interactions, the root-mean-square scattering angle appears
changed and dependent on spin orientation with respect to vectors
$\vec{N_m}$, $\vec{N_T}$ and $\vec{n}$.
When measuring MDM and $ T_{odd} $ spin rotation in a bent
crystal, one can eliminate parity violating rotation by the
following way (see Fig.\ref{fig5}).


\begin{figure}[htbp]
    \centerline{\includegraphics[width=10 cm]{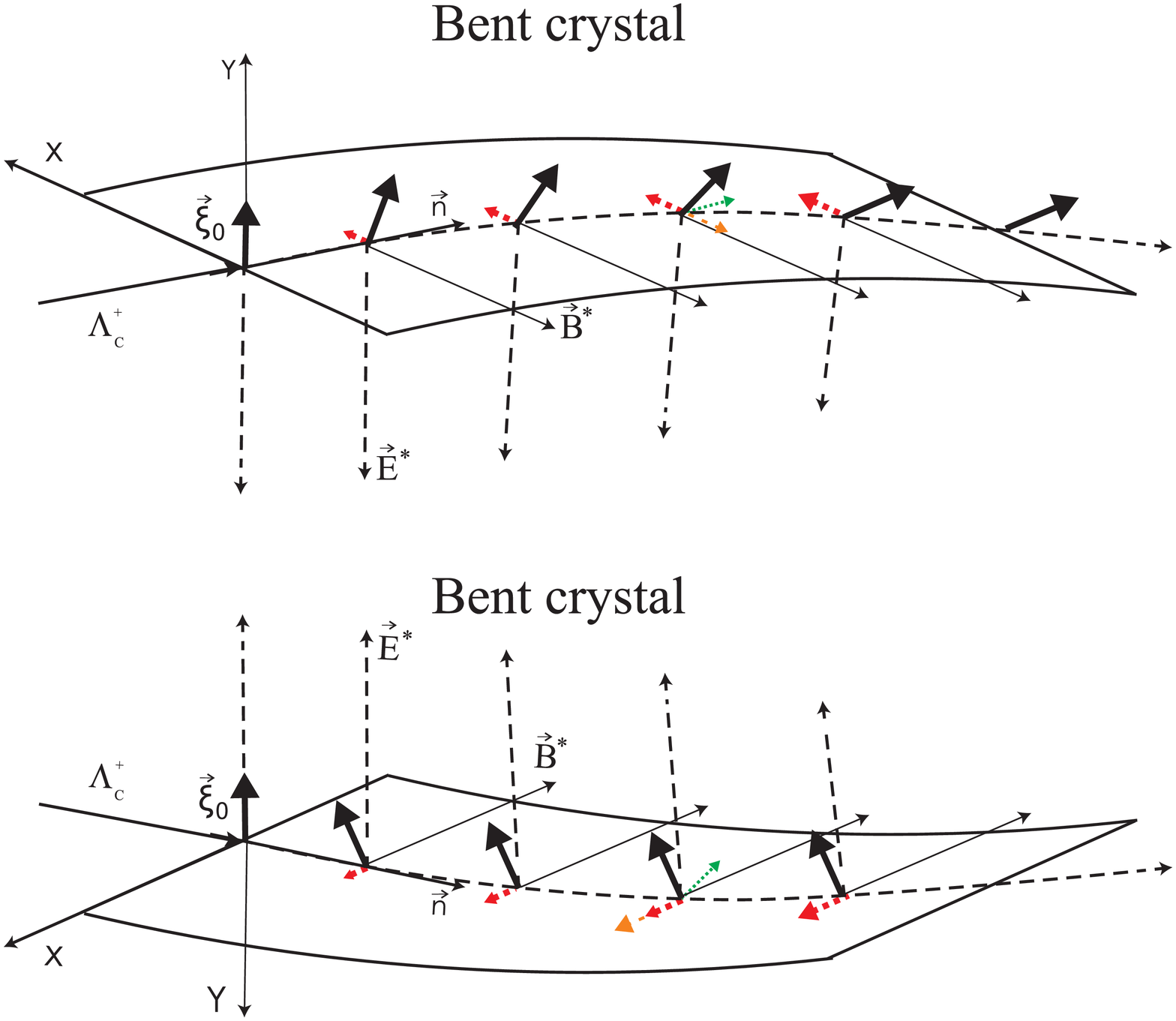}}
    \caption{By turning the crystal $180^{\circ}$ around the direction of incident baryon momentum
One could observe that $P_{odd} $ spin rotation does not change, while the sign of MDM and  $T_{odd} $ spin
rotations does due to change of the electric field direction. Subtracting results of measurements for two opposite crystal positions one could obtain the angle of rotation, which does not depend on $P_{odd} $ effect.
    }
    \label{fig5}
\end{figure}


\begin{figure}[htbp]
     \centerline{\includegraphics[width=10 cm]{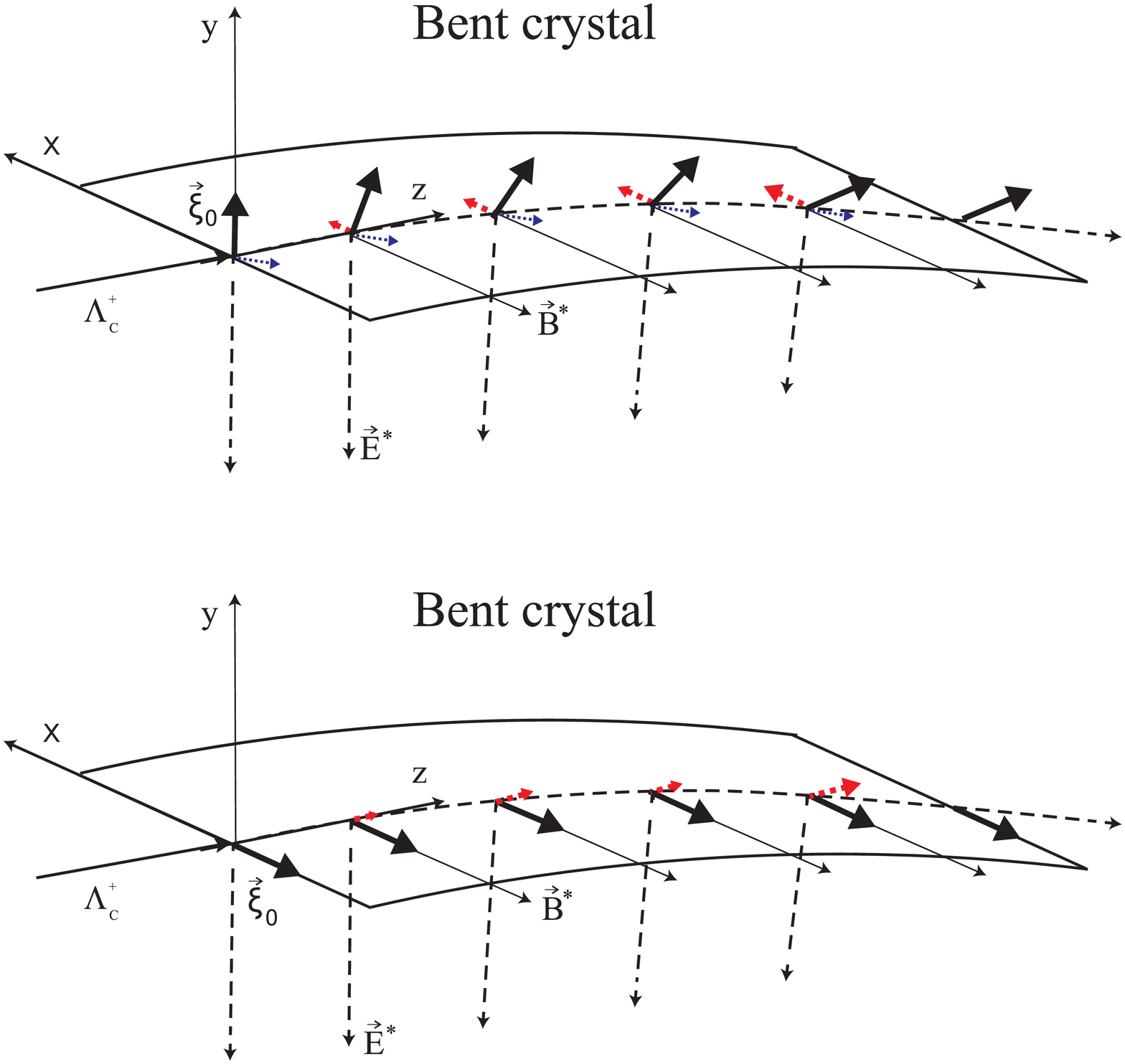}}
    \caption{Separation of the contributions caused by MDM and $T_{odd} $ spin rotation is
possible when comparing experimental results for two initial orientations of polarization vector $\xi$. Namely:$ \vec{\xi} \parallel \vec{N_m}$ and $ \vec{\xi} \parallel \vec{N_t}$, i.e. the initial $\xi$ is
parallel to the bending axis of the crystal or $\vec{E}$ .
In real situation rotating the crystal by $ 90 \circ$ so that direction of $ S_0 $ is parallel to $ B* $ can be more convenient.}
    \label{fig6}
\end{figure}

By turning the crystal $ 180^{\circ} $ around the direction of
incident baryon momentum one could observe that $ P_{odd} $ spin
rotation does not change, while the sign of MDM and $ T_{odd} $ spin
rotations does due to change of the electric field direction.
Subtracting results of measurements for two opposite crystal
positions one could obtain the angle of rotation, which does not
depend on $ P_{odd} $ effect.
Such measurement also can be made using the idea presented in
\cite{b10}, according to which to control systematic uncertainties
two crystals (up and down bending) should be used to induse
opposite spin precession to channeled baryons.
Separation of the contributions caused by MDM and $ T_{odd} $ spin
rotation is possible when comparing experimental results for two
initial orientations of polarization vector $ \vec{\xi} $.
Namely: $\vec{\xi} \parallel \vec{N_m} $ and $\vec{\xi} \parallel
\vec{N_T} $, i.e. the initial $ \vec{\xi}  $ is parallel to the
bending axis of the crystal or $\vec{E}$ (see Fig.\ref{fig6}).

\section{Conclusion}
Besides electromagnetic interaction the channelled particle, which
moves in a crystal, experiences weak interaction with electrons
and nuclei, as well as strong interaction with nuclei.
Measurements of polarization vector and angular distribution of
particles scattered by axes (planes) of unbent crystal enable to
obtain limits for the EDM value and for values of constants
describing P- and T-odd interactions.
The same measurements also allow studying magnetic dipole moment
of charged and neutral particles.
Investigation of left-right asymmetry by the use of two unbent
crystals makes it possible to measure EDM, MDM and other constants
without studying the angular distribution of decay products of
scattered particles:
 it is
sufficient to measure the angular distribution of intensity of
flow of particles experienced double scattering.
When analyzing particle's spin rotation, which is caused by
electric dipole moment interaction with electric field, one should
consider non-invariant spin rotations both $P_{odd}$, $T_{even}$
and $P_{odd}$, $T_{odd}$, resulting from weak interaction with
electrons and nuclei.
As demonstrated hereinabove, spin precession of channelled
particles in bent crystals at the LHC gives unique possibility for
measurement of constants determining $T_{odd}$, $ P_{odd}$ (CP)
violating interactions and $P_{odd}$, $T_{even}$ interactions of
baryons with electrons and nucleus (nucleons), similarly to the
possibility of measuring electric and magnetic moments of charm,
beauty and strange charged baryons.
%
For a particle moving in a bent crystal a new effect, which is
caused by nonelastic processes, arises: in addition to the spin
precession around three directions $ \vec{N}_m,\vec{N}_T,\vec{n}$
the spin dichroism effect causes the appearance of the spin
components in directions of $ \vec{N}_m,\vec{N}_T,\vec{n}$.
%
To separate P-noninvariant rotation from the MDM- and EDM-induced
($T_{odd}$) spin rotations both the method of turning crystal by $
180 ^{\circ} $ and the method of using two crystals suggested in
\cite{b10} can be used. To separate contributions caused by MDM
and $T_{odd}$ interactions, two bent crystals placed perpendicular
to each other can be used.

\section{Appendices}

\subsection{Scattering amplitude} \label{ssec1}

    As follows from (\ref{eq12}):
    \begin{equation}
    F_{j}(0)= f_{j}(0)-i\frac{k}{4 \pi}\int f_{j}^*(\vec{k}''
    -\vec{k}')f_{j}(\vec{k}'' -\vec{k}) d\Omega_{ k''}. \label{eq13}
    \end{equation}
    The integral in (\ref{eq13}) is {equal to} the total cross-section
    of the elastic coherent scattering by a nucleus (atom). According
    to the optical theorem:
    \begin{equation}
    Im f_{j}(0)= \frac{k}{4 \pi} \sigma_{tot}=\frac{k}{4 \pi}
    \sigma_{elast} + \frac{k}{4 \pi} \sigma_{nonelast}. \label{eq14}
    \end{equation}
    In  contrast to the matter with chaotically distributed scatterers
    the amplitude $ F_j (0) $ in crystal
    is expressed as follows:
    \begin{equation}
    F_{j}(0)=\tilde{f}_{j}(0), \tilde{f}_{j}(0)=
    f_{j}(0)-\frac{k}{4\pi} \sigma_{elast}. \label{eq15}
    \end{equation}
    In other words, the cross-section of elastic coherent scattering
    in crystal does not contribute to the imaginary part of the
    amplitude $F_j(0)$.
    This imaginary part is solely determined by the cross-section of
    nonelastic processes (cross-section of reactions):
    \begin{equation}
    F_{j}(0)=Re F_{j}(0)+ i Im F_{j}(0) = Re F_{j}(0)+i\frac{k}{4 \pi}
    \sigma_{nonelast}. \label{eq16}
    \end{equation}
The nonzero-angle scattering possesses the similar features.
{This fact becomes clear when one uses the equality, which is
    correct for elastic scattering} \cite{bn14}:
\begin{equation}
Im f_{elast} (\vec{k}' -\vec{k})= \frac{k}{4 \pi} \int
f_{elast}^*(\vec{k}'' -\vec{k}')f_{elast}(\vec{k}'' -\vec{k})
d\Omega_{ k''}. \label{eq17}
\end{equation}
and subtracts the elastic scattering contribution from the
imaginary part of $ f(\vec{k}' -\vec{k})$ using (\ref{eq12}).

The same result can be obtained by considering the interaction
    with the scatterer in terms of the perturbation theory.
In case when the first Born approximation is used, scattering
    amplitude $f^{(1)}(\vec{k}'-\vec{k})$ has zero imaginary part:

\begin{equation}
Im f^{(1)}_{aa} (\vec{k}' -\vec{k})=0. \label{eq18}
\end{equation}
The non-zero imaginary part arises, when one uses the second order
Born approximation.

Lets remind that T-operator, which determines the scattering
amplitude (see (\ref{eq5b})), satisfies the following equation
\cite{bn13, bn15}:

\begin{equation}
T=V+V\frac{1}{E-H_{0}+i\eta}T. \label{eq19}
\end{equation}
where $V$ is the interaction energy, $H_{0}$ is the Hamilton
operator of colliding systems located at large distance from each
other.

As a result for the  elastic coherent scattering amplitude$
f_{aa}$ with the accuracy up to the second order terms over the
interaction energy, one gets:

\begin{eqnarray}
f_{aa} (\vec{k}' -\vec{k})= -(2 \pi)^2 \frac{m \gamma}{\hbar^2} (<\Phi_{\vec{k}'a}| V |\Phi_{\vec{k}a}> + \nonumber\\
 + <\Phi_{\vec{k}'a}| V \frac{1}{E_{a}
(\vec{k})-H_{0}+i \eta} V |\Phi_{\vec{k}a}>), \label{eq20}
\end{eqnarray}
where $ \Phi_{\vec{k}a} $ is an eigenfunction of Hamilton operator $ H_{0}$,
\begin{equation}
\Phi_{\vec{k}a}=\frac{1}{(2 \pi)^{3/2}} e^{i \vec{k}
    \vec{r}}\Phi_{a}, \label{eq200}
\end{equation}
$\Phi_a $ is the wave function of scatterer stationary states and
$H_{0}\Phi_{\vec{k}a}=E_{a}(\vec{k})\Phi_{\vec{k}a}$.
Using the completeness of function $\Phi_{\vec{k}a} $ and
replacing "1" in (\ref{eq20}) by
$\sum_{\vec{k}''b}|\Phi_{\vec{k}''b}><\Phi_{\vec{k}''b}|=1$ one
obtains the sum over the intermediate states $b$, which includes
states with $ b=a $. This term contains the following expression:
\begin{equation}
\frac{1}{E_{a}(\vec{k})-E_{a}(\vec{k}'')+i\eta}= P
\frac{1}{E_{a}(\vec{k})-E_{a}(\vec{k}'')} - i \pi \delta
(E_{a}(\vec{k})-E_{a}(\vec{k}'')). \label{eq21}
\end{equation}
The $"P"$ symbol in the real part in
{(\ref{eq21})}
means, that all integrals containing the  $"P"$ symbol in
(\ref{eq20}) are the principal-value integrals.

This real part contribution is small as compared to the first Born
approximation, therefore, it will not be further considered.
The imaginary unit in the second term of (\ref{eq21}), which is
proportional to the $ \delta $ function, leads to occurrence of an
imaginary part in amplitude $ f (\vec{k}' -\vec{k}) $.
{Substitution of the expression with $ \delta$-function into
    (\ref{eq20}) makes it obvious that the term in the sum, in which
    $b=a$, is equivalent to the term subtracted from the amplitude  $
    f_{aa} (\vec{k}' -\vec{k}) $ in (\ref{eq12}).}
As a result only terms caused by nonelastic processes and
reactions with $b\neq a$  make contributions to the imaginary part
of the amplitude in crystal.
{To simplify consideration below, when expressing $F$, the
    contribution from the elastic coherent scattering to the imaginary
    part of the amplitude $f_{aa}$ will not be considered and the
    second term in (\ref{eq12})  will not be also written explicitly.}

\subsection{Effective potential energy of a spin-particle moving
    close to the crystal planes (axes)} \label{ssec2}

Let's consider the expression for  effective potential energy in
detail. According to (\ref{eq172},\ref{eq25},\ref{eq26}) the
contributions to the effective potential energy are caused by
interactions of different types  including short-range and
long-range interactions.
In the presence of several types of interaction, to describe their
different contributions to the scattering amplitude, it is
convenient to separate scattering caused only by long-range
interactions and present amplitude in following form:
\begin{equation}
f(\vec{q})= f_{long} (\vec{q})+ f_{short long} (\vec{q}),
\label{eq27}
\end{equation}
where $f_{long} (\vec{q})$ is a scattering amplitude determined by
long-range coulomb and magnetic interactions  (assuming that
short-range interactions are absent), $f_{short long} (\vec{q})$
is a scattering amplitude determined by short-range interactions
(calculating this amplitude waves scattered by long-range
interactions were used as an incident waves). For general
scattering theory in the presence of several interactions see, for
example, \cite{bn13,bn15}.

When several types  of interactions influence  on the scattering
amplitude, it can be easily studied with the help of perturbation
theory. Let interaction energy $V$ be a sum of several
interactions: $V=\sum_i V_i$. Then at the first Born approximation
scattering amplitude is a sum of scattering amplitudes caused by
every interaction separately: $f=\sum_i f_1 (V_i)$. But in the
second Born approximation additional term $ f_2 $, which is
determined by the following expression, appears in the scattering
amplitude (see \cite{bn13,bn14,bn15})
\begin{equation}
f_2 = V \frac{1}{E-H_0 -i\eta}V = \sum_p V_p \frac{1}{E-H_0
    -i\eta} \sum_l V_l , \label{eq28}
\end{equation}
As one can see, equality (\ref{eq28}) contains interference
contributions to $f$ proportional to $V_p V_l$.

Let us now consider how different terms, which amplitude
(\ref{eq182a}) includes, contribute to the effective potential
energy of particle interaction with the crystal.

The Coulomb amplitude, described by the first term in
(\ref{eq182a}), leads to conventional expression for potential
energy of interaction between a charged particle and  a plane
(axis).

The second term $A_s (\vec{q}) $ is caused by the short-range
interaction. Amplitude $A_s (\vec{q})$ can be written as:

\begin{equation}
A_{s}(q)=A_{nuc} (q)\Phi_{osc} (\vec{q}), \label{eq29}
\end{equation}
where $A_{nuc}(q)$ is the spin independent part of the amplitude
of elastic scattering by the resting nucleus, $\Phi_{osc}
(\vec{q})$ is the form-factor caused by nucleus oscillations in
crystal.

Owing to the short-range kind of strong interactions amplitude,
$A_{nuc} (\vec{q})$ is equal to zero-angle scattering amplitude
$A(0)$ within the range of scattering angles  $\vartheta\leq
\frac{1}{kR_{osc}}\ll 1 $.

Form-factor $\Phi_{osc}(\vec{q})$ has the form \cite{bn14}:
\begin{equation}
\Phi_{osc}(\vec{q})=\sum_n \rho_n <\varphi_n (r)| e^{-i
    \vec{q}
    \vec{r}} |\varphi_n (r)>=
\int e^{-i \vec{q} \vec{r}} N_{nuc} (\vec{r}) d^3 r, \label{eq30}
\end{equation}
where $\varphi_n (r) $ is the wave function describing vibrational
state of nuclei in crystal, summation $\sum_{n}{\rho_n}$ means
statistical averaging with Gibbs distribution over vibrational
states of nucleus in crystal. Let's remind, that squared
form-factor $\Phi_{osc} (\vec{q}) $ is equal to Debye-Waller
factor, $ N_{nuc} (\vec{r}) $ is a probability density of
vibrating nuclei detection in point $\vec{r}$, $\int N_{nuc}
(\vec{r}) d^3r = 1 $.
As a result, equations (\ref{eq31}), (\ref{eq322})  can be obtained.

\subsection{Effective potential energy determined by the anomalous
    magnetic moment} \label{ssec3}

According to (\ref{eq182a}) the scattering amplitude, which is
determined by baryon's anomalous magnetic moment, has the form:
\begin{equation}
\hat{F}_{magn} (q)= B_{magn} (q) \vec{\sigma} [\vec{n} \times
\vec{q}]. \label{eq33}
\end{equation}

Defining the scattering amplitude at the first step one could
solely consider magnetic scattering and its interference with
Coulomb scattering (see (\ref{eq27})), and at the second step add
the term caused by interference between magnetic and nuclear
interactions.

For the first step perturbation theory can be used. In the first
order of perturbation theory the interference of the magnetic
moment scattering by the Coulomb field with the Coulomb scattering
of baryon electric charge by the Coulomb field is absent. The
amplitude $ \hat{F}^{(1)}$ reads as:
\begin{equation}
\hat{F}_{magn}^{(1)} (\vec{q}) = i f_{coul}
(\vec{q})\frac{\hbar}{m c} (\frac{g-2}{2})\frac{1}{2} \vec{\sigma}
[\vec{n} \times \vec{q}] , \label{eq34}
\end{equation}
where  $f_{coul} (\vec{q})$ is the amplitude of coulomb scattering
of a baryon by an atom in the first Born approximation; $
\vec{n}=\frac{\vec{k}}{k}$, $m$ is the baryon mass.

It should be noted that the coefficient, by which  $\vec{\sigma}$
is multiplied, in the expression for amplitude
$\hat{F}_{magn}^{(1)} (\vec{q})$ is purely imaginary. After
substitution of (\ref{eq34}) into (\ref{eq25}) and summation over
$\tau_x$ one obtains the expression for effective interaction
energy as follows:
\begin{equation}
\hat{U}_{magn}(x) = -\frac{e \hbar}{2 m c} \frac{g-2}{2} \vec{\sigma} [\vec{E}_{plane}(x) \times \vec{n}] ,
\label{eq35}
\end{equation}
where $\vec{E}_{plane}(x)$ denotes the electric field, produced by
the crystallographic plane in point $x$. In axis case
$U_{magn}(\vec{\rho})$ can be obtained by replacement of $x$ by
$\vec{\rho}$ in (\ref{eq35}) and $\vec{E}_{plane} (x)$ by
$\vec{E}_{axis} (\vec{\rho})$, respectively.

Using (\ref{eq35}) and Heisenberg equations for spin operator, the
motion equation for  polarization vector (\ref{eq1}, \ref{eq2})
for the case of  $ B=0 $ and $\gamma  \gg 1 $ can be obtained.

Effective interaction energy (\ref{eq35}) can be rewritten as
follows:

\begin{equation}
\hat{U}_{magn} = -\frac{e \hbar}{2 m c} \frac{g-2}{2}E_{x
    plane}(x) \vec{\sigma} \vec{N}, \label{eq36}
\end{equation}
where $\vec{N}=[\vec{n}_x \times \vec{n}]$ is the unit vector,
$\vec{n}_x \perp \vec{n}$, unit vector $\vec{n}$ is parallel to
the crystallographic plane.

Expression (\ref{eq36}) for the effective potential energy
comprises factor, which is purely real.
However, the coefficient in the expression for scattering
amplitude $ \hat{F}(\vec{q})$, by which  $\vec{\sigma}$ is
multiplied, has non-zero both  real and imaginary parts. Due to
this fact, the effective potential energy  $\hat{U}$ also has
non-zero both real and imaginary parts.

In the second order of perturbation theory this coefficient in
amplitude $ \hat{F}(\vec{q})$ is not purely imaginary as well --
it has a non-zero real part. By means of (\ref{eq12}),
(\ref{eq20})-(\ref{eq21}) the following expression for the
contribution $\tilde{F}^{(2)} (q)$ to the amplitude $
\hat{F}(\vec{q})$ can be obtained:

\scriptspace=-1.5pt

\begin{eqnarray}
\label{eq37} && \tilde{F}^{(2)} (\vec{q}=\vec{\tau}) =i \frac{k}{4
    \pi\hbar^2 c^2} \times  \nonumber \\
& & \times \left\{ <\Phi_a|\iint e^{-i \vec{\tau}
    \vec{r}_{\perp}}
\left[ \int \hat{V} (\vec{r}_{\perp},z)dz \right]^2 d^2 r_{\perp} |\Phi_a>  - \right. \nonumber \\
& & \left. -  \iint e^{-i \vec{\tau} \vec{r}_{\perp}} \left[ \int
<\Phi_a| \hat{V}(\vec{r}_{\perp},z) | \Phi_a> dz \right]^2 d^2
r_{\perp} \right\} = \nonumber \\
& = & i \frac{k}{4 \pi \hbar^2 c^2} \iint e^{-i \vec{\tau}
    \vec{r}_{\perp}} \left\{ <\Phi_a| \left[ \int \hat{V}
(\vec{r}_{\perp},z)dz \right]^2 |\Phi_a> - \right.  \nonumber \\
& & \left. -  \left[ \int <\Phi_a| \hat{V} (\vec{r}_{\perp},z)
|\Phi_a> dz \right]^2 \right\} d^2 r_{\perp}
= \\
& = & i \frac{k}{4 \pi \hbar^2 c^2} \iint   e^{-i \vec{\tau}
    \vec{r}_{\perp}} \left\{
{\overline { \left[ \int \hat{V}(\vec{r}_{\perp},z)dz
        \right]^2} } - \right. \nonumber \\
&&- \left.  \Bigg[ \overline{\int \hat{V}(\vec{r}_{\perp},z)dz} \,
\Bigg]^2 \right\} d^2 r_{\perp}\,, \nonumber
\end{eqnarray}
where $ \hat{V}(\vec{r}_{\perp},z)=
\hat{V}_{coul}\,(\vec{r}_{\perp},z)+\hat{V}_{magn}\,(\vec{r}_{\perp},z)$,
$\hat{V}_{magn}\,(\vec{r}_{\perp},z)=- \mu_a \,  \vec{\sigma}
[\vec{E} (\vec{r}_{\perp}, z) \times \vec{n}]$, $z$ axis of the
coordinate system is directed along the unit vector $\vec{n}$,
$\vec{n}$ is the unit vector directed along the particle momentum
before scattering  $\hbar k $, $\mu_{a}$ is the anomalous magnetic
moment of the particle $\mu_{a}=\frac{e \hbar }{2 m c}
(\frac{g-2}{2})$.

\scriptspace=0pt

When deriving (\ref{eq37}), it was considered that the particle
energy is much greater than the electrons' binding energy in atoms
and the atoms' binding energy in crystal.
As a result it is possible at first to examine scattering by
electrons and nuclei, which rest in points $r_i$,  and then to
average the result over the electrons and nuclei positions with
wave functions $|\Phi_a>$ (impulse approximation, for example see
\cite{bn13}). The overline in (\ref{eq37}) and hereinafter denotes
such kind of averaging.
The contribution caused by interference between magnetic and
nuclear scattering, and  the contributions determined by the
particle squared magnetic moment should complete the expression
mentioned above. For positively charged particles, moving far from
the top of the potential barrier, the contribution caused by
interactions with nuclei is suppressed and will be omitted in
consideration hereinafter.
Contributions proportional to the particle squared magnetic moment
are smaller then those caused by interference between magnetic and
Coulomb scattering and will, thus,  be also omitted.
After substitution of (\ref{eq37}) into (\ref{eq25}) and summation
over $\tau_x$ the following expression for the contribution to the
effective potential energy caused by the amplitude $
\tilde{F}_{magn} (\vec{\tau})$ can be obtained:
\begin{equation}
\hat{U}^{(2)}_{magn}(x)= -i \frac{1}{4 d_y d_z m c^2}
(\frac{g-2}{2}) \frac{\partial}{\partial x} \overline{\delta V^2
    (x)} \vec{\sigma}\vec{N}, \label{eq38}
\end{equation}
where $\vec{N}=[\vec{n}_{x} \times \vec{n}]$, $\vec{n}_{x} \perp
\vec{n}, \vec{n}_{x} $ is the unit vector along axis $x$,
$$ \overline {\delta V^2 (x)}\!=\!\int\!\left\{\overline{ \left[\int V_{coul} (x,y,z) dz \right]^2}\!-\!\left[ \overline { \int V_{coul} (x,y,z) dz} \right]^2 \right\}\!dy
$$

Similarly for the case of axis it can be obtained:
\begin{equation}
\hat{U}^{(2)}_{magn}(\vec{\rho})= -i \frac{1}{d_z m c^2}
(\frac{g-2}{2}) \vec{\sigma} [\nabla_{\rho} \overline{\delta V^2
    (\vec{\rho})} \times \vec{n}]. \label{eq39}
\end{equation}
For the axisymmetric case:
\begin{equation}
\hat{U}^{(2)}_{magn}(\rho)= -i \frac{1}{4 d_z m c^2}
(\frac{g-2}{2}) \frac{\partial}{\partial \rho}\overline{\delta V^2
    (\rho)} [\vec{n}_{\rho}  \times \vec{n}], \label{eq40}
\end{equation}

\noindent where $\overline{\delta V^2 (\vec{\rho})} =  \overline{
    \left[ \int V_{coul} (\vec{\rho},z)dz \right]^2}  -
\left[\overline{\int V_{coul} (\vec{\rho},z)dz}\right]^2$,
$\vec{n}_{\rho}=\frac{\vec{\rho}}{\rho}$ is the unit vector,
$\vec{n}_{\rho} \perp \vec{n}$.

In the planar channeling case $\hat{U}_{magn}$  is determined by the expression (\ref{eq41}).

\subsection{Effective potential energy $\hat{U}$ determined by
    spin-orbit interaction} \label{ssec4}

According to (\ref{eq182a}) the part of the scattering amplitude
caused by strong spin-orbit interaction has the form:
\begin{equation}
\hat{F}_{s sp-orb} (\vec{q}=\vec{\tau})=B_s (\vec{\tau})
\vec{\sigma}[\vec{n} \times \vec{\tau}]. \label{eq42}
\end{equation}
The coefficient $B_s (\vec{\tau}) $ can be expressed similar to
(\ref{eq29}) as follows:
\begin{equation}
B_s (\vec{\tau})=B_{ nuc} (\vec{\tau})\Phi_{osc} (\vec{\tau}),
\label{eq43}
\end{equation}
where $B_{s nuc} (\vec{\tau})$ describes scattering by a resting
nucleus, $\Phi_{osc}(\vec{\tau})$ is the form-factor determined by
nucleus oscillations in crystal.

In the considered case, similar to the approach used when deriving
(\ref{eq31}), the short-range character of the nuclear forces and
small (as compared to the amplitude of nucleus oscillations)
nucleus radius  enables assumption $B_{s nuc} (\vec{\tau})\approx
B_{s nuc} (0)$. It is important that the coefficient $B_{s nuc}
(0)$ has non-zero both real and imaginary parts:
\begin{equation}
B_{s nuc} (0)=B_{s nuc}' + i B_{s nuc}''. \label{eq44}
\end{equation}
This is similar to the case of amplitude, which describes
scattering of the magnetic moment by the atom (nucleus).
To obtain the expression for the effective potential energy the
summation over  $\tau_x$ should be conducted in (\ref{eq25}). The
resulted expression is similar to that for $\hat{U}_{magn}$. For
example, for the crystal plane case see expression (\ref{eq46}) for $ \hat{U}_{ssp-orb} $.
Let us remind that the contribution determined by elastic
scattering, which is described by the second term in (\ref{eq13}),
is negligibly small in comparison with nonelastic contributions to
the amplitude and, therefore, can be omitted.

\subsection{Effective potential energy $\hat{U}$ determined by
    $ P_{odd} $ and $ T_{even} $ interactions} \label{ssec5}

The next group of terms, which are proportional to $B_w$, is
determined by weak $ P_{odd} $  and $ T_{even} $ interactions. According to
(\ref{eq182a}) the corresponding terms in the scattering amplitude
can be written as:
\begin{equation}
\hat{F}_w (\vec{q})=(B_{we}(\vec{q}) + B_{w nuc}(\vec{q}))
\vec{\sigma}\vec{N}_{w}\,. \label{eq600}
\end{equation}
Contribution $B_{we}(\vec{q})$ caused by parity violating weak
interaction between the baryon and electrons can be expressed as
follows:

\begin{equation}
B_{we}(\vec{q})= \tilde{B}_{we}(\vec{q})\Phi_e(\vec{q}),
\label{eq601}
\end{equation}
where $\tilde{B}_{we}$ is the coefficient defining baryon elastic
scattering amplitude by resting electron
$\hat{f}_{we}(q)=\tilde{B}_{we}\vec{\sigma}\vec{N}_{w}$,
$\Phi_e(\vec{q})=\int e^{-i \vec{q} \vec{r}} N_e (\vec{r})d^3r$,
$\int N_e (\vec{r})d^3r=Z$, $Z$ is the nucleus charge.
Minor corrections caused by thermal oscillations of atoms' centers
of gravity will not be considered below. To take them into
consideration one should multiply $\Phi_e(\vec{q})$ by
$\Phi_{osc}(\vec{q})$, which is the form-factor defined by
oscillations of atoms nucleus.

Term $B_{w nuc}(\vec{q})$ (see (\ref{eq600})), which is caused by
parity violating weak interaction between the baryon and nucleus,
reads as follows:
\begin{equation}
B_{w nuc}(\vec{q})= \tilde{B}_{w nuc}(\vec{q})\Phi_{osc}(\vec{q}),
\label{eq602}
\end{equation}
where $\tilde{B}_{w nuc}$ is the coefficient defining the
amplitude of a baryon elastic scattering  by a resting nucleus
$\hat{f}_{w nuc}=\tilde{B}_{w nuc}\vec{\sigma}\vec{N}_w$\,.

Due to the short-range character of P-violating interactions, when
angle $\vartheta \simeq \frac{\tau}{k}\ll 1 $, coefficients
$\tilde{B}_{we}(\vec{q})\simeq \tilde{B}_{we}(0)$ and
$\tilde{B}_{w nuc}(\vec{q})\simeq \tilde{B}_{w nuc}(0)$ .
As a result  expressions for the effective potential energy $
\hat{U}_w $ of P-violating interaction of a baryon with a crystal
plane (axis) were obtained (see (\ref{eq606}),(\ref{eq605})).

\subsection{Effective potential energy $\hat{U}$ determined by the
    electric dipole moment and other T-nonivariant interactions}
\label{ssec6}
Let us consider now the electric dipole moment and other
T-nonivariant contributions to the spin rotation. According to
(\ref{eq182a}) the corresponding terms in the scattering amplitude
can be written as:
\begin{equation}
\hat{F}_T (q)=( B_{EDM}(q) + B_{Te}(q) + B_{T nuc}(q))
\vec{\sigma}\vec{q}. \label{eq47}
\end{equation}
Let's consider the term $\hat{F}_{EDM}
(q)=B_{EDM}(\vec{q})\vec{\sigma}\vec{q}$.
The coefficient $B_{EDM}(q)$ has non-zero both real and imaginary
parts $B_{EDM}(q)=B_{EDM}'+iB_{EDM}''$.
By the approach used for deriving $\hat{F}_{magn} (q)$, for
$\hat{F}_{EDM }(\vec{q})$ one can obtain:
\begin{eqnarray}
&  \hat{F}&\!_{EDM} (\vec{q})= -i\frac{m \gamma d}{2 \pi \hbar^2} V_{coul}(\vec{q})\vec{\sigma}\vec{q}+ \\
&+&\frac{k}{4 \pi \hbar^2 c^2} \times \nonumber \\
& \times\!\! & \iint\!e^{-i \vec{q}_{\perp} \vec{r}_{\perp}
}\left\{\overline{\left[\int \hat{V}
    (\vec{r}_{\perp},z)dz\right]^2}\!
-\!\left[\int\!\overline{ \hat{V}\!(\vec{r}_{\perp},z)}dz\right]^2\!\right\}\!d^2 r_{\perp},\nonumber\\
\nonumber \label{eq48}
\end{eqnarray}
where $\hat{V}(\vec{r})=V_{coul}(\vec{r})+V_{EDM}(\vec{r}) $,
$V_{EDM}=-D\vec{\sigma}\vec{E}$ is the energy of interaction
between the electric dipole moment $D$ and the electric field
$\vec{E}$,  $D=ed$,  $e$ is the electric charge of the particle.
Using (\ref{eq25}), expression (\ref{eq49})  for the potential
energy of interaction between the particle and the crystal plane
can be obtained.
Let's remind that amplitude $\hat{F}_{T} (\vec{q})$ contains terms
both caused by EDM and determined by short-range T- noninvariant
interactions between the baryon and electrons and nuclei $B_{T
    e}(\vec{q})$ and $B_{T nuc} (q)$.
Contributions caused by these terms should also be added to the
effective potential energy of the interaction between the baryon
and nuclei of the crystal $\hat{U}_{T} (x)$:

\begin{equation}
\hat{U}_{T} (x)= \hat{U}_{EDM} + \hat{U}_{Te} + \hat{U}_{T nuc} =
-(\alpha_{T}(x) + i\delta_{T}(x))\vec{\sigma}\vec{N}_{T},
\label{eq52}
\end{equation}
where $\alpha_{T} =\alpha_{EDM}+ \alpha_{Te}+ \alpha_{T nuc},
\delta_{T}=\delta_{EDM}+ \delta_{Te}+ \delta_{T nuc}$.

Expressions for coefficients  $\alpha_{Te(nuc)}$ and
$\delta_{Te(nuc)}$ can be evaluated in terms of scattering
amplitude by the following way. Let's define the form-factor
determined by electrons distribution in atom and nucleus
oscillations.
\begin{equation}
B_{Te} (\vec{q})=\tilde{B}_{Te}(\vec{q})\Phi_{e}(\vec{q}),
B_{Tnuc} (\vec{q})=\tilde{B}_{Tnuc}(\vec{q})\Phi_{osc}(\vec{q}),
\label{eq53}
\end{equation}
where $\Phi_{e}(\vec{q})=\int e^{-i \vec{q} \vec{r}} N_{e}
(\vec{r}) d^3 r$, $ N_{e} (\vec{r}) $  is electrons distribution
density in atom, $\int N_{e}(\vec{r})d^3 r=Z$, $Z$ is the nucleus
charge, $\Phi_{osc}(\vec{q})$ is determined by (\ref{eq30}),
$\tilde{B}_{Te}$ is the coefficient defining amplitude of baryon
scattering by resting electron
$\hat{f}_{Te}=\tilde{B}_{Te}(\vec{q})\vec{\sigma}\vec{q} $,
$ \tilde{B}_{nuc} (q)$  is the coefficient defining amplitude of
baryon scattering  by resting nucleus
$\hat{f}_{Tnuc}=\tilde{B}_{nuc}(\vec{q})\vec{\sigma}\vec{q}$.
Let's remind that in compliance with (\ref{eq12}) the contribution
caused by elastic coherent scattering should be subtracted from
the amplitude $B_{T}$. However, at high energies this contribution
is negligibly small in comparison with nonelastic contributions to
the amplitude and, therefore, can be omitted.

Due to the short-range character of T-noninvariant interactions at
angle $\vartheta\simeq\frac{\tau}{k} \ll 1$ coefficients
$\tilde{B}_{Te}(\vec{q})\simeq\tilde{B}{_{Te}}(0)$ and
$\tilde{B}_{nuc}(\vec{q})\simeq\tilde{B}_{nuc}(0)$ .

Due to the short-range character of T-noninvariant interactions at
angle $\vartheta\simeq\frac{\tau}{k}<<1$ coefficients
$\tilde{B}_{Te}(\vec{q})\simeq\tilde{B}{_{Te}}(0)$ and
$\tilde{B}_{nuc}(\vec{q})\simeq\tilde{B}_{nuc}(0)$ . As a result,
the following expressions can be obtained:
\begin{eqnarray}
&\hat{U}_{Te}(x)&=i \frac{2 \pi \hbar^2}{m \gamma d_y d_z} \tilde{B}_{Te} (0) \frac{dN_{e} (x)}{dx} \vec{\sigma}\vec{N}_{T},\nonumber\\
&\hat{U}_{Tnuc}(x)&=i \frac{2 \pi \hbar^2}{m \gamma d_y d_z} \tilde{B}_{Tnuc} (0) \frac{dN_{nuc} (x)}{dx} \vec{\sigma}\vec{N}_{T},\nonumber\\
&N_{e(nuc)}(x)&=\int N_{e(nuc)}(x,y,z)dydz, \label{eq54}
\end{eqnarray}
Coefficients $\tilde{B}_{Te} (0)$ and $\tilde{B}_{Tnuc} (0)$ are
complex values:
\begin{equation}
\tilde{B}_{Te(nuc)}(0)=\tilde{B}'_{Te(nuc)}+i\tilde{B}''_{Te(nuc)}.\nonumber
\end{equation}
As a result expression (\ref{eq50}) can be obtained.

\begin{acknowledgments}
Author would like to thank Dr. Nicola Neri for fruitful
discussions.
\end{acknowledgments}

 \end{document}